\documentclass[twoside,twocolumn,9pt]{article}
\usepackage{extsizes}
\usepackage[super,sort&compress,comma]{natbib} 
\usepackage[left=1.5cm, right=1.5cm, top=1.785cm, bottom=2.0cm]{geometry}
\usepackage{balance}
\usepackage{mathptmx}
\usepackage{sectsty}
\usepackage{graphicx} 
\usepackage{lastpage}
\usepackage{verbatim}
\usepackage[format=plain,justification=justified,singlelinecheck=false,font={stretch=1.125,small,sf},labelfont=bf,labelsep=space]{caption}
\usepackage{float}
\usepackage{fancyhdr}
\usepackage{fnpos}
\usepackage[english]{babel}
\addto{\captionsenglish}{%
  
}
\usepackage{array}
\usepackage{droidsans}
\usepackage{charter}
\usepackage[T1]{fontenc}
\usepackage[usenames,dvipsnames]{xcolor}
\usepackage{setspace}
\usepackage[compact]{titlesec}
\usepackage{hyperref}
\usepackage{booktabs}

\usepackage{siunitx}
\usepackage{isotope}
\usepackage{longtable}
\usepackage{fixltx2e}
\usepackage{float,graphicx,amsmath,multirow}
\usepackage{color}
\usepackage{booktabs}
\usepackage{gensymb}
\usepackage{booktabs}
\usepackage{makecell}
\usepackage{tabularx}
\usepackage{ragged2e}
\usepackage[version=4]{mhchem}
\definecolor{cream}{RGB}{222,217,201}

\begin{document}

\pagestyle{fancy}
\thispagestyle{plain}
\fancypagestyle{plain}{
\renewcommand{\headrulewidth}{0pt}
}

\makeFNbottom
\makeatletter
\renewcommand\LARGE{\@setfontsize\LARGE{15pt}{17}}
\renewcommand\Large{\@setfontsize\Large{12pt}{14}}
\renewcommand\large{\@setfontsize\large{10pt}{12}}
\renewcommand\footnotesize{\@setfontsize\footnotesize{7pt}{10}}
\makeatother

\renewcommand{\thefootnote}{\fnsymbol{footnote}}
\renewcommand\footnoterule{\vspace*{1pt}%
\color{cream}\hrule width 3.5in height 0.4pt \color{black}\vspace*{5pt}} 
\setcounter{secnumdepth}{5}

\makeatletter 
\renewcommand\@biblabel[1]{#1}            
\renewcommand\@makefntext[1]%
{\noindent\makebox[0pt][r]{\@thefnmark\,}#1}
\makeatother 
\renewcommand{\figurename}{\small{Fig.}~}
\sectionfont{\sffamily\Large}
\subsectionfont{\normalsize}
\subsubsectionfont{\bf}
\setstretch{1.125} 
\setlength{\skip\footins}{0.8cm}
\setlength{\footnotesep}{0.25cm}
\setlength{\jot}{10pt}
\titlespacing*{\section}{0pt}{4pt}{4pt}
\titlespacing*{\subsection}{0pt}{15pt}{1pt}

\fancyfoot{}
\fancyfoot[LO,RE]{\vspace{-7.1pt}\includegraphics[height=9pt]{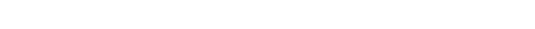}}
\fancyfoot[CO]{\vspace{-7.1pt}\hspace{13.2cm}\includegraphics{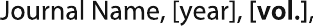}}
\fancyfoot[CE]{\vspace{-7.2pt}\hspace{-14.2cm}\includegraphics{head_foot/RF}}
\fancyfoot[RO]{\footnotesize{\sffamily{1--\pageref{LastPage} ~\textbar  \hspace{2pt}\thepage}}}
\fancyfoot[LE]{\footnotesize{\sffamily{\thepage~\textbar\hspace{3.45cm} 1--\pageref{LastPage}}}}
\fancyhead{}
\renewcommand{\headrulewidth}{0pt} 
\renewcommand{\footrulewidth}{0pt}
\setlength{\arrayrulewidth}{1pt}
\setlength{\columnsep}{6.5mm}
\setlength\bibsep{1pt}                                                                                                                      

\makeatletter 
\newlength{\figrulesep} 
\setlength{\figrulesep}{0.5\textfloatsep} 

\newcommand{\topfigrule}{\vspace*{-1pt}%
\noindent{\color{cream}\rule[-\figrulesep]{\columnwidth}{1.5pt}} }

\newcommand{\botfigrule}{\vspace*{-2pt}%
\noindent{\color{cream}\rule[\figrulesep]{\columnwidth}{1.5pt}} }

\newcommand{\dblfigrule}{\vspace*{-1pt}%
\noindent{\color{cream}\rule[-\figrulesep]{\textwidth}{1.5pt}} }

\makeatother

\twocolumn[
  \begin{@twocolumnfalse}
{\includegraphics[height=30pt]{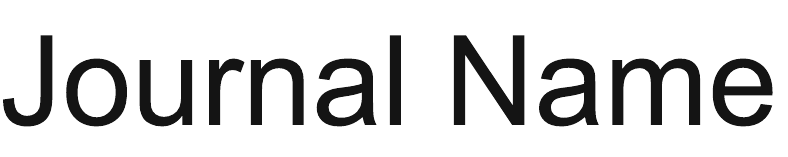}\hfill\raisebox{0pt}[0pt][0pt]{\includegraphics[height=55pt]{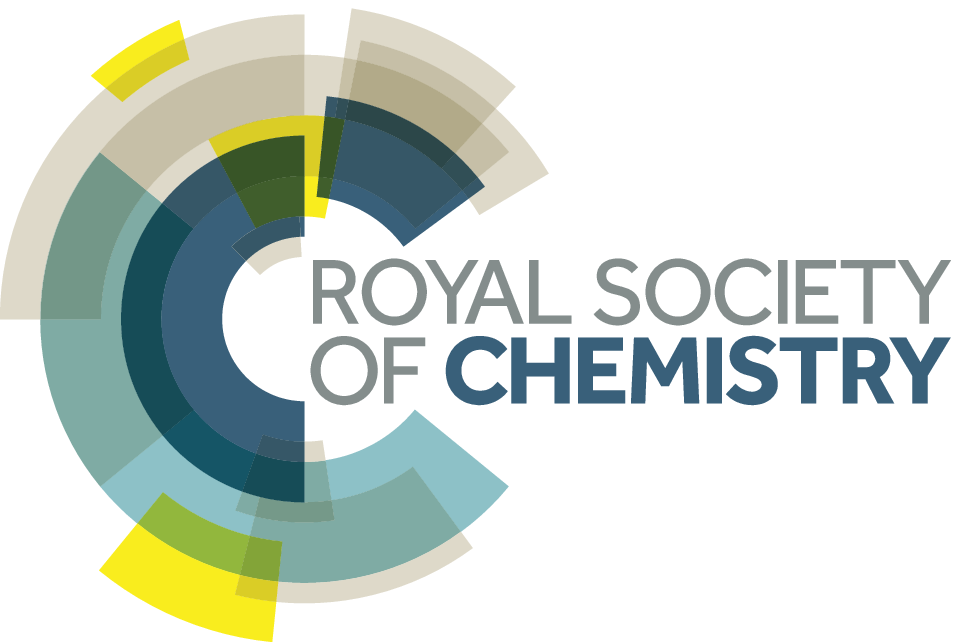}}\\[1ex]
\includegraphics[width=18.5cm]{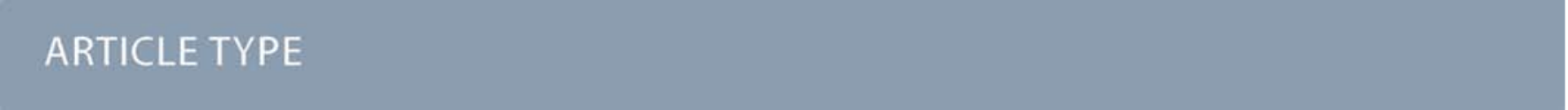}}\par
\vspace{1em}
\sffamily
\begin{tabular}{m{4.5cm} p{13.5cm} }

\includegraphics{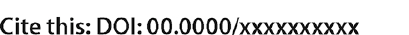} & \noindent\LARGE{\textbf{Implementation of Rare Isotopologues into Machine Learning of the Chemical Inventory of the Solar-Type Protostellar Source IRAS 16293-2422}} \\
\vspace{0.3cm} & \vspace{0.3cm} \\

 & \noindent\large{Zachary T.P. Fried,\textit{$^{a}$} Kin Long Kelvin Lee,\textit{$^{b}$} Alex N. Byrne,\textit{$^{c}$} and Brett A. McGuire\textit{$^{d,e}$}} \\

\includegraphics{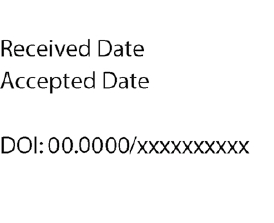} & \noindent\normalsize{Machine learning techniques have been previously used to model and predict column densities in the TMC-1 dark molecular cloud.  In interstellar sources further along the path of star formation, such as those where a protostar itself has been formed, the chemistry is known to be drastically different from that of largely quiescent dark clouds. To that end, we have tested the ability of various machine learning models to fit the column densities of the molecules detected in source B of the Class 0 protostellar system IRAS 16293-2422.  By including a simple encoding of isotopic composition in our molecular feature vectors, we also examine for the first time how well these models can replicate the isotopic ratios. Finally, we report the predicted column densities of the chemically relevant molecules that may be excellent targets for radioastronomical detection in IRAS 16293-2422B.} \\

\end{tabular}

 \end{@twocolumnfalse} \vspace{0.6cm}
]


\renewcommand*\rmdefault{bch}\normalfont\upshape
\rmfamily
\section*{}
\vspace{-1cm}


\footnotetext{\textit{$^{a}$~Department of Chemistry, Massachusetts Institute of Technology, Cambridge, MA 02139; E-mail: zfried@mit.edu}}
\footnotetext{\textit{$^{b}$~Accelerated Computing Systems and Graphics Group, Intel Corporation, 2111 NE 25th Ave., Hillsboro, OR 97124}}
\footnotetext{\textit{$^{c}$~Department of Chemistry, Massachusetts Institute of Technology, Cambridge, MA 02139}}
\footnotetext{\textit{$^{d}$~Department of Chemistry, Massachusetts Institute of Technology, Cambridge, MA 02139; E-mail: brettmc@mit.edu}}
\footnotetext{\textit{$^{e}$~National Radio Astronomy Observatory, Charlottesville, VA 22903}}

\footnotetext{\dag~Electronic Supplementary Information (ESI) available: [details of any supplementary information available should be included here]. See DOI: 00.0000/00000000.}

\section{Introduction}
\label{sec:intro}

The observation of interstellar molecules is a central component of astrochemical studies. Molecular species have shaped our understanding of star \citep{van98} and planet formation \citep{obe11}, can trace stellar outflows \citep{bac01}, interstellar shocks \citep{sch97}, and protoplanetary disks \citep{gin19}, and can serve as probes of the physical conditions of interstellar sources such as the temperature \citep{ho79}. However, until recently, in order to model interstellar abundances and predict new molecules for detection, observations have relied on complex chemical models based on a vast network of interconnected reactions (e.g. \citet{rua16}, \citet{wak15}). While these astrochemical models can be excellent tools to explore specific chemical processes that occur in space, their predictive ability can also be quite limited for several reasons (e.g. \citet{mcg15}).  Firstly, these models are by definition incomplete representations of the true chemical complexity of the interstellar medium because network expansions rely on human input. Additionally, the networks are oftentimes dependent on uncertain extrapolated rate constants\citep{heb09}. 

In an attempt to predict molecular abundances without the need for complete networks, \citet{lee21} introduced a novel methodology involving machine learning. A major benefit of their approach contrasts traditional astrochemical modeling, as it requires no prior knowledge of the conditions of an interstellar source or any reaction pathways involving the previously detected molecules. Instead, abundances are expressed purely in terms of a chemical vector space. Simple regression algorithms were shown to significantly outperform traditional astrochemical models in reproducing the abundances of molecules already observed, and provided a straightforward way to extrapolate to yet undetected molecules. 

An interstellar source for which this machine learning technique could be effectively applied is the Class 0 protostar IRAS 16293-2422B (hereafter referred to as IRAS 16293B). IRAS 16293B is one component of the protostellar system IRAS 16293, which is located in the L1689 region of the $\rho$ Ophiuchus cloud complex.  Interferometric observations initially revealed two protostellar sources in IRAS 16293 (source A and source B), separated by around 5.1$^{\prime\prime}$ \citep{woo89,mun92,loo00}. Further high-resolution studies then confirmed that source A is in fact composed of two compact sources (source A1 and A2), making IRAS 16293 a triple protostellar system \citep{mau20}. Extensive observations have been made of IRAS 16293 with the Atacama Large Millimeter/submillimeter Array (ALMA) as part of the Protostellar Interferometric Line Survey (PILS) program \citep{jor16}. The submillimeter spectrum toward source B is especially rich with more than 10,000 features detected \citep{jor16}. The line widths of the spectral peaks are also extremely narrow for a star forming region ($\sim$1 km/s FWHM), which significantly reduces line confusion and makes this an excellent source for molecular detections. 

The predictive power of the machine learning method introduced by \citet{lee21} may be especially useful for IRAS 16293B since a large portion of the molecular lines in the interstellar line survey remain unassigned. In fact, as of 2018, \citet{taq18_o2} noted that approximately 70\% of the 5$\sigma$ transitions identified in the ALMA Band 6 dataset were unassigned. If successful, this method might be able to provide an unbiased list of astrochemical targets not yet detected but which might be abundant enough to be contributing to the unassigned molecular lines.  If subsequently detected, these molecules and their abundances could then be used to further constrain both the machine learning model and traditional network-based astrochemical models of low mass protostars.  These models provide invaluable insight into the chemical processes and conditions relevant to the formation of stars similar to our Sun.

One aspect of interstellar chemistry that was not treated in the work of \citet{lee21} was the incorporation of isotopically substituted species. While certainly such rare isotopologues are present and detectable in TMC-1 \citep{bur_iso}, detections of these species are more common toward IRAS 16293B and therefore provide substantially more insight into the chemical and physical history of this source. Therefore, it is desirable to update the machine learning model to include isotopologues. Molecules in IRAS 16293B consistently display isotopic ratios that are enhanced compared to the mean solar value and other interstellar sources, especially deuterium (D) and \isotope[13]{C} substituted species. In fact, the deuterated isotopologues of ethanol, ketene, acetaldehyde, formic acid, formamide, and isocyanic acid were all first detected toward this source \citep{jor18, cou16}. Various doubly and triply deuterated species have been detected as well (e.g. \citet{ily22, per18}). Additionally, the \isotope[12]{C}/\isotope[13]{C} ratios of dimethyl ether, methyl formate, ethanol, and glycolaldehyde toward IRAS 16293B are all much lower than the \isotope[12]{C}/\isotope[13]{C} ratio of the local ISM \citep{jor16, jor18}. By convention, the deuterium ratios are reported as D/H while the \isotope[13]{C} ratios are reported in the inverse manner. Therefore, a high D/H ratio and low \isotope[12]{C}/\isotope[13]{C} ratio both denote isotopic enhancement.

A large portion of the remaining unassigned spectral peaks are predicted to arise from isotopically substituted species. In fact, \citet{jor16} note that only 25\% of the transitions correspond to the most common organic molecules detected in hot cores, including formaldehyde, methanol, methyl cyanide, isocyanic acid, ethanol, acetaldehyde, methyl formate, dimethyl ether, and ketene. Following this, they predict that the majority of the remaining transitions are likely related to various isotopically substituted molecules as well as more complex organic species. Thus, it is also vital to accurately model the column densities of isotopically substituted molecules so that the high abundance isotopologues in this source can be predicted, measured as needed in the laboratory, and their signals in the PILS survey identified and assigned.  

Additionally, machine learning predictions of isotopic ratios are also useful since these ratios can act as tracers of the evolutionary history of an interstellar source along with the conditions, timescales, and pathways of molecular formation. For example, deuterium fractionation relies on gas-phase isotope exchange reactions that are strongly dependent on the temperature. Consequently, the deuterium fraction is a tracer of the conditions of the interstellar environment during molecular formation, with a high D/H ratio (i.e. high deuterium fraction) indicating cold formation temperatures \citep{wat76,mil89}. Therefore, accurate prediction of isotopic ratios would allow us to gain insight into the details of molecular formation and source history without requiring a dedicated search for these isotopically-substituted species that are often present in fairly low abundance.

In this work we apply the machine learning technique introduced by \citet{lee21} to IRAS 16293B. The machine learning pipeline used for this project along with the isotopic encoding is described in Section~\ref{sec:model}. Section~\ref{sec:regression} then presents the ability of the supervised machine learning regressors to model the molecular column densities in this source. Using these trained regression models, we obtain an unbiased list of predicted high-abundance targets for astronomical observation. Analysis of these molecular targets is presented in Section~\ref{sec:targets}. Next, in Section~\ref{sec:isotopes} we test the ability of the regressors to model the isotopic ratios in this source. Finally, a list of high predicted column density isotopologues is provided in Section \ref{sec:iso_predict}.

\section{Dataset}
\label{sec:dataset}

The molecules included in the dataset for this work were mostly detected through observations from the PILS survey \citep{jor16}. The line widths of the spectral peaks toward IRAS 16293A are much wider than those observed toward source B \citep{jor16}. Consequently, there is much more overlap of the spectral peaks, making the identification of individual signals more challenging and resulting in fewer definitive molecular detections toward source A. Our analysis therefore focused solely on IRAS 16293B. Most molecules were detected at a one-beam (0.5$^{\prime\prime}$) offset position from the continuum peak of source B in the south-west direction. The coordinates of this one-beam offset position are $\alpha$\textsubscript{J2000} = 16\textsuperscript{h}32\textsuperscript{m}22\textsuperscript{s}.58, $\delta$\textsubscript{J2000} = $-24^{\circ}28^{\prime}32.8^{\prime\prime}$. A few of the species, however, were detected at a half-beam offset. For these molecules, the column densities have been reduced by a factor of 2.136 to account for this different pointing position \citep{dro19}. In total, our dataset includes 98 molecules. Of these, 43 are main isotopologues and 55 are isotopically substituted species. All molecules in the dataset are listed in Table \ref{appendixTable} in the Appendix. Our dataset contains 27 deuterium substituted species, 15 \isotope[13]{C} substituted species, two \isotope[15]{N} substituted species, four \isotope[34]{S} substituted species, two \isotope[33]{S} substituted species, two \isotope[17]{O} substituted species, and three \isotope[18]{O} substituted species. Additionally, several doubly deuterated molecules along with one triply deuterated molecule is included.

 \section{Model Description}
 \label{sec:model}

The machine learning pipeline has three main components: (1) molecular featurization, (2) modeling the column densities in IRAS 16293B using supervised regressors, and (3) prediction of high abundance astrochemical targets. An outline of this entire process is depicted in Figure \ref{schematic} in the Appendix. In this work, the process of molecular featurization and regression is quite similar to the methods introduced by \citet{lee21}. In summary, a \textsc{Mol2vec}\citep{jae18} model is first trained on a dataset of 3,634,046 molecules collected from various online databases like Pubchem \citep{pubchem}, ZINC \citep{zinc}, and the NASA PAH database \citep{nasa1, nasa2, nasa3}. This trained embedding model then creates 70 dimensional feature vectors for all molecules in the dataset, in addition to the detected interstellar species. 

In order to include isotopologues in the training set and investigate isotopic ratios, it was necessary to encode isotopic composition in the feature vectors. In its current form, \textsc{Mol2vec} is not able to fully capture isotopic information. For example, it creates unique vectors for deuterium-substituted molecules but not for molecules that are substituted with \isotope[13]{C}. This is because the molecular substructures are first encoded using Morgan fingerprints\citep{mor65}, which do not by-default capture differences in \isotope[13]{C}-substituted isotopologues (e.g. the default \textsc{RDKIT}-constructed \citep{rdkit} Morgan fingerprint  of \ce{H2CO} and \ce{H2}\ce{^{13}CO} are identical).  To differentiate between each of the isotopologues in our dataset, we ensure that the \textsc{Mol2vec}-generated vectors are identical for all isotopologues of the same species and then add 19 extra dimensions that encode isotopic information as well as the chemical environment of the isotopic substitution. The isotopic encoding is designed as follows:

\begin{itemize}
\item Dimensions 1-9: Number of D, \isotope[34]{S}, \isotope[33]{S}, \isotope[36]{S}, \isotope[13]{C}, \isotope[17]{O}, \isotope[18]{O}, \isotope[15]{N}, and \isotope[37]{Cl} atoms in the molecule.
\item Dimensions 10-12: Whether the \isotope[13]{C} atoms are sp, sp\textsuperscript{2}, or sp\textsuperscript{3} hybridized.
\item Dimension 13: Whether the substituted \isotope[13]{C} atom is bonded to oxygen.
\item Dimension 14-15: Whether a deuterium atom is bonded to carbon or oxygen.
\item Dimension 16: Number of non-hydrogen atoms in the molecule.
\item Dimensions 17-18: Whether there is an oxygen or carbon atom two bonds away from the substituted deuterium.
\item Dimension 19: Number of deuterium atoms bonded to carbonyl carbons.
\end{itemize}

\begin{figure}[h!]
\begin{center}
\includegraphics[width=\columnwidth]{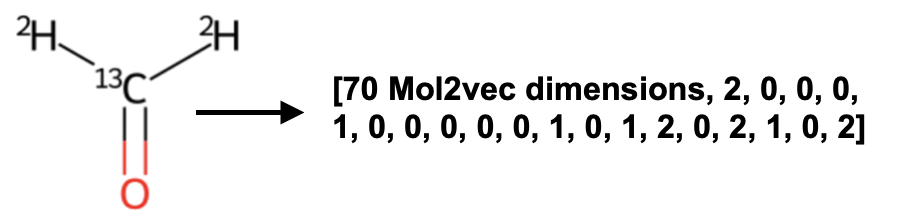}
\caption{Depiction of how isotopic composition is encoded in the molecular feature vectors.}\label{encoding}
\end{center}
\end{figure}

The \textsc{RDKit} module was used to obtain hybridization and bonding information \citep{rdkit}. Because of the limited number of unique isotopologues in our dataset, the selection of these hand-picked features was largely dependent on which chemical substructures are present in enough molecules to constitute a reasonably sized training set. More specifically, each of the isotopic features denoted by dimensions 10-19 are present in at least three molecules in the dataset. Each of the selected features also has a notable impact on the average isotopic ratio. 

Additionally, \textsc{Mol2vec} was unable to differentiate between several conformers of the detected molecules. Examples include ethylene glycol (for which both the aGg' and gGg' conformers were detected) as well as monodeuterated \ce{CH2DCH2OH} and \ce{CH2DOCH3}. For consistency, we inputted the column density of the most stable or abundant conformer in each case.

Using the resulting feature vectors as inputs and the $\log_{10}$ column densities as outputs, the data was split 80/20 into training and testing sets. In order to mitigate data leakage, all isotopologues of the same molecule were assigned to either the training or testing set. The datapoints were then bootstrapped with Gaussian noise in order to increase the effective dataset size to 800 and control overfitting.

These resulting training and testing sets were then fed into two separate supervised machine learning regressors: Gaussian process regression (GPR) \citep{gpr} and Bayesian ridge regression (BR). These models learn relationships between the vector components to map the molecular features to the column density data. Each of the models were implemented with the \textsc{scikit-learn} Python module \citep{scikit-learn}. We determined the optimal hyperparameters for each model by first splitting the data into traning and testing sets and then running a 5-fold grid search on the training data.

GPR is a nonparametric model that defines a probability distribution over all functions that can map the molecular descriptors to the column densities. It is therefore able to handle nonlinear relationships in the data. A kernel provides the model with prior knowledge regarding the shape and smoothness of the functions. Similarly to \citet{lee21}, the kernel we used was a linear combination of the rational quadratic, dot product, and white noise kernels. Along with the kernel function, additional hyperparameters include a noise value added to the kernel matrix diagonal that denotes the inherent Gaussian noise of the training observations. 

BR is a linear regressor that takes a probabilistic approach to optimize the ridge regression model coefficients.  It does this by using a Gamma distribution prior for the regularization coefficients. These parameters are then optimized through maximization of the log marginal likelihood. For this regressor, the hyperparameters define the shape and inverse scale of the Gamma distribution priors over the various model parameters.

Similarly to \citet{lee21}, a linear model is included in order to provide a baseline performance using an extremely simple model with a limited number of parameters. Bayesian ridge was specifically chosen due to its ability to report prediction uncertainties, which allows us to gauge the confidence level of the predicted values. A GPR model then displays the ability of a more complex model to improve upon this baseline regressor. GPR was also chosen due to its probabilistic and nonlinear nature. The reported uncertainties are fairly informative and interpretable since they can be linked directly to the designed covariance matrix. 

Following the training of the regression models, the column densities of the molecules that are most chemically similar to those detected in IRAS 16293B were predicted using the trained models. K-means clustering with k = 10 was used to cluster the entire dataset of 3,634,046 feature vectors. Each of the molecules detected in IRAS 16293B were assigned to a single cluster. Thus, we only considered the molecules assigned to this cluster when searching for detectable new species. 

When analyzing the ability of the regressors to model the molecules in the training set and subsequently predict the column densities of the species in the testing set, we were limited to the molecules that have been previously detected toward IRAS 16293B. We were therefore only able to gauge the performance of these models on molecules that are relevant to this fairly small and homogeneous dataset. Consequently,  we proceeded to remove the molecules that had no or few chemically similar examples in the dataset since the models did not have sufficient training examples to learn the required relationships for these species and we had no ability to gauge the accuracy of the model predictions. For this additional filtering, we removed molecules that contained atoms other than hydrogen, carbon, sulfur, nitrogen, and oxygen because all of the previously detected molecules were predominantly composed of these atoms. Additionally, we also removed the remaining free radical species because nitrous oxide is the only free radical for which a column density has been derived toward IRAS 16293B. This minuscule training set of free radicals resulted in the models not being able to sufficiently handle these molecules. For example, the free radical counter-parts of various molecules had much higher predicted column densities than the observed values of the parent species. This is very unlikely due to the instability of free radicals and the general under abundance of radicals in protostellar sources. In fact, while 36.5\% of interstellar molecules were first detected in star-forming regions, only 13.5\% of radical species were first detected in these protostellar sources \citep{mcg21}. This under abundance may be due to the larger gas-phase chemical inventory and warmer kinetic temperatures leading to a greater number of destruction partners for these highly reactive species. Following these filtering steps, the column densities of the remaining 84,863 molecules were then predicted using the previously trained regression models.

\section{Results and Discussion}

\subsection{Regression Analysis}
\label{sec:regression}

\begin{figure*}
\begin{center}

\includegraphics[width=\textwidth]{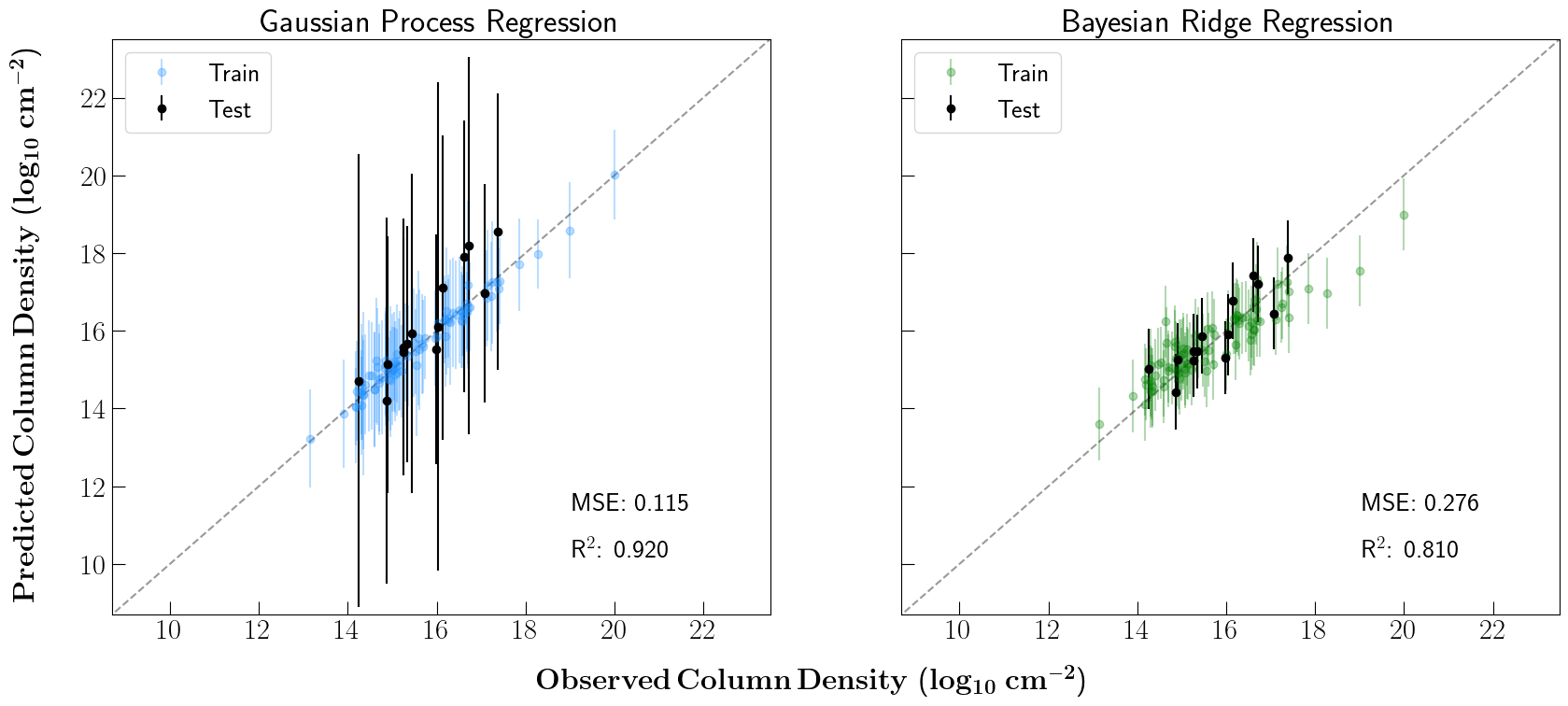}
\caption{Training and testing results of each supervised regression model. An 80/20 train/test split was used. The reported MSE and R\textsuperscript{2} are for the combined training and testing data. 1$\sigma$ uncertainties are shown.}\label{fig1}
\end{center}
\end{figure*}

Figure \ref{fig1} shows the training and testing results of each regression model. The train/test splits were consistent in each case. Just as shown in \citet{lee21}'s study of TMC-1, each of the regressors were able to accurately model the column densities in this source. The strong performance on the test set also provides confidence that the models can generalize well to relevant molecules that were not included in the training set.

Additionally, although the BR model showed an ability to precisely model the column densities with lower uncertainties than the GPR regressor, it was mainly limited by its linear mapping. With our current isotope encoding, a linear model will be unable to fully capture the relevant isotopic fractionation. For example, the difference between a singly and doubly deuterated molecule is in-part denoted with a 2 instead of a 1 in a single vector dimension. That said, the difference in the column densities of singly and doubly deuterated species is typically not simply 2/1 and can differ significantly between molecules. Thus, for the remainder of the analysis, a GPR model was used since a nonlinear mapping was required. 

The large error bars on the GPR predictions are in part due to the small size of the dataset. Additional molecular detections (especially of main isotopologues) toward this source will allow for further constrained predictions. Moreover, despite the overall strong performance of the GPR model, this regressor overpredicts ethylene glycol (\ce{OHCH2CH2OH}) and dimethyl ether (\ce{CH3OCH3}) by over one order of magnitude. This prediction inaccuracy is likely because these molecules have few nearby neighbors in the training set. In fact, these molecules are the 10\textsuperscript{th} and 12\textsuperscript{th} furthest species from any neighbor in the dataset, respectively. Ethylene glycol is especially unique in that it is the only molecule in the dataset containing two hydroxyl groups. Additionally, ethylene glycol’s nearest neighbors are methoxymethanol and ethanol, each of which are more abundant.

Another notable prediction error is the slight overprediction of chloromethane and underprediction of methanol since it highlights the shortcomings of our molecular featurization. \textsc{Mol2vec} creates molecular feature vectors by combining vector representations of chemical substructures. Therefore, small molecules with some shared substructures have extremely similar feature vectors. In this case, \textsc{Mol2vec} generates similar feature vectors for all molecules that contain a methyl group bonded to a single heteroatom. Despite obvious chemical differences, the resulting vector representations of chloromethane and methanol are therefore very similar. Since methanol is one of the most abundant molecules in the dataset and chloromethane is one of the least abundant, the resulting prediction errors are fairly unsurprising.

In order to test the efficacy of the chosen kernel for the GPR model, we predicted the column density of cimetidine \ce{C10H16N6S}. This molecule is far more complex than any other species in our dataset and would certainly have an extremely low abundance in the interstellar medium. Therefore, an effective kernel would also produce a low column density prediction for this species. Ultimately, the trained GPR model predicted this molecule to have a column density of \num{6.94e6} \si{\per\centi\meter\squared}, which is nearly eight orders of magnitude lower than any main isotopologue in the dataset. This provides confidence that the model is not over-fitting to the dataset and simply learning to predict each column density to be in the range of the detected species. 

\subsection{Targets for astrochemical study}
\label{sec:targets}

\begin{figure}[h]
\includegraphics[width=\columnwidth]{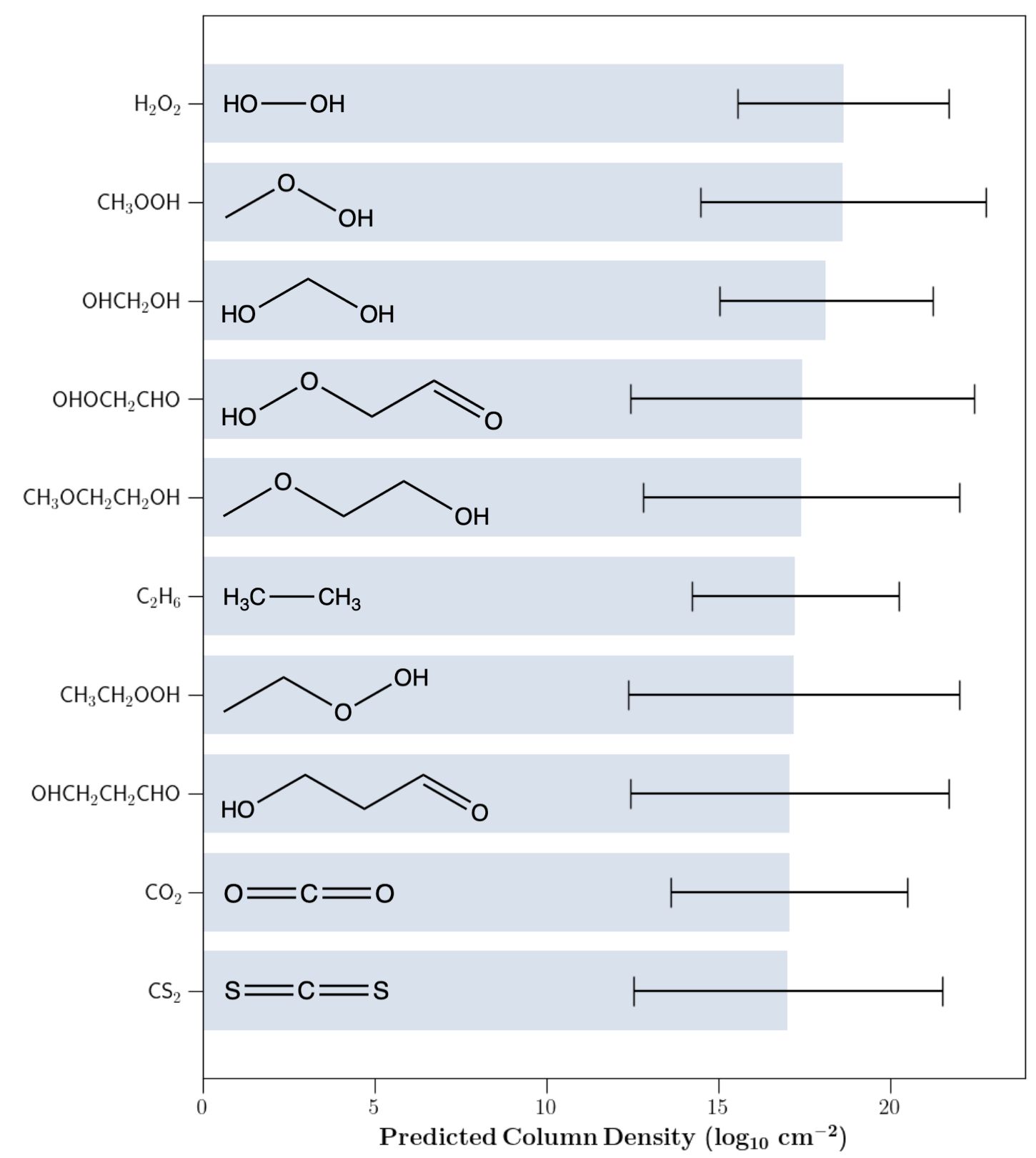}
\caption{The 10 undetected molecules with the highest predicted column densities. The molecules were drawn from the collection of species that were assigned to the same cluster as each of the detected molecules through k-means clustering. Predictions were made using the trained Gaussian process regression model. 1$\sigma$ column density uncertainties are shown. }\label{fig4}
\end{figure}

Using the trained GPR model, we then predicted the column densities of the aforementioned 84,863 astrochemically relevant molecules assigned to the same cluster as the IRAS 16293B detections. To enhance the confidence in our predicted values and further limit our investigation to chemically relevant molecules, our analysis was solely focused on species for which the 1$\sigma$ prediction uncertainty was less than five orders of magnitude; there were 242 such species. Figure \ref{fig4} shows the 10 molecules with the highest predicted column densities. All predictions are provided in the associated \href{https://github.com/zfried/IRAS_ML_Predictions}{GitHub repository}.

\begin{table*}

\begin{center}
\caption{\label{tablePredictions} Average chemical composition of the 20 highest and lowest predicted abundance astrochemical targets.}
\label{table1}
\begin{tabular}{c|p{1.5in}p{1.5in}} 
\hline

 & \Centering{20 Highest Predicted Abundance Molecules}  & \Centering{20 Lowest Predicted Abundance Molecules} \\ [1ex] 
 \hline

 Mean \texttt{\#} of Oxygen Atoms & \Centering{1.65} & \Centering{0.75} \\ 
 
 Mean \texttt{\#} of Nitrogen Atoms & \Centering{0.10} & \Centering{1.05} \\
 
 Mean \texttt{\#} of Sulfur Atoms & \Centering{0.10} & \Centering{0.75} \\

 Mean Degree of Unsaturation & \Centering{0.75} & \Centering{1.65}  \\

 Mean \texttt{\#} of Heavy Atoms & \Centering{3.85} & \Centering{4.25}  \\

 Mean Molecular Weight (amu) & \Centering{59.66} & \Centering{74.35}\\
 [0.5ex] 
 \hline

\end{tabular}
\end{center}
\end{table*}

The chemical composition of the predicted molecules is displayed in Table \ref{tablePredictions}. Oxygenated hydrocarbons were typically predicted to be in high abundance while those containing nitrogen and sulfur were predicted to have lower column densities. Of the 20 highest predicted column density molecules, 15 contain at least one oxygen atom, while 2 contain a nitrogen atom, and 1 contains a sulfur atom. The preference for oxygen-substituted molecules is not surprising since the most abundant detected species in IRAS 16293B are carbon monoxide (CO), methanol (\ce{CH3OH}), formaldehyde (\ce{H2CO}), methyl formate (\ce{HCOOCH3}), dimethyl ether (\ce{CH3OCH3}), carbonyl sulfide (\ce{OCS}), and ethanol (\ce{CH3CH2OH}) -- each of which contain an oxygen atom. 

Highly saturated molecules were also predicted to be very abundant in IRAS 16293B. This is to be expected in a protostellar source since hydrogenation is very efficient on grain surfaces. Therefore, many of the species that are sublimated from grains as the protostar heats the surroundings are highly saturated (e.g. \citealt{lin15, fed15, woo02, gar08}).

The predictions also display a preference for lighter molecules that contain less heavy atoms. This also matches the detected chemical inventory in IRAS 16293B, in which the seven highest abundance molecules each contain four or less heavy atoms. 

The proceeding subsections highlight the astrochemical relevance of some of the molecules with the highest predicted column densities in Figure \ref{fig4}. It is important to note that while a high column density is beneficial for interstellar molecular detectability, various additional factors must also be considered including the magnitude of the dipole moment, the spectral pattern, and intrinsic line strengths.

\subsubsection{Hydrogen Peroxide (\ce{H2O2})} 
Hydrogen peroxide was first observed toward the SM1 core in $\rho$ Oph A through several torsion-rotation transitions \citep{ber11}. This molecule is proposed to form on grain surfaces via successive hydrogen additions to \ce{O2} \citep{du12}. If hydrogen peroxide were to be detected, another possible molecular candidate is \ce{HO2}. This radical is generated after the first of these successive hydrogen additions to molecular oxygen and was also detected toward $\rho$ Oph A with a similar abundance to hydrogen peroxide \citep{par12}. 

\subsubsection{Methyl Hydroperoxide (\ce{CH3OOH})}
Methyl hydroperoxide is a very attractive candidate for interstellar detection. This organic peroxide is structurally similar to many of the small chemical species that have been detected in the interstellar medium, such as hydrogen peroxide \citep{ber11}. Additionally, multiple energetically feasible products of methyl hydroperoxide UV photodissociation have been detected in space, including the OH and \ce{CH3O} radicals \citep{the93, mcn65, cer12}. The microwave and millimeter-wave spectra of this molecule have been previously experimentally studied and assigned, thus allowing for radioastronomical detection \citep{tyb92}. To the best of our knowledge, no dedicated search for this molecule has been conducted toward any interstellar source. 

\subsubsection{Methanediol (\ce{OHCH2OH})}
Methanediol is the simplest diol molecule and is of astrochemical interest due to its similarity to previously detected species such as methanol. Through modeling efforts, it has been proposed that this molecule is generated via grain surface reactions of the OH and \ce{CH2OH} radicals \citep{gar08}. For years, methanediol has been extensively studied in the aqueous phase (e.g. \citet{moh87, mat80, rya02}). However, gaseous production and detection of this molecule was only reported in the past year \citep{zhu22}. To our knowledge, there have been no previous high-resolution microwave studies of this species. Thus, without experimental rotational parameters, definitive interstellar detection using radioastronomy is currently impossible. 

\subsubsection{Methoxyethanol (\ce{CH3OCH2CH2OH})}
This molecules is in the same chemical family as methoxymethanol (\ce{CH3OCH2OH}) and methoxyethane (\ce{CH3OCH2CH3}), which have large column densities of \num{1.4e17} \si{\per\centi\meter\squared} and \num{1.8e16} \si{\per\centi\meter\squared} toward IRAS 16293B, respectively \citep{man20}. The methoxy radical (\ce{CH3O}) has been detected in space and is proposed to form through methanol photodissociation, gas-phase reactions between the OH radical and methanol, and hydrogen addition to \ce{H2CO} \citep{ch3o}. Additionally, methoxymethanol has been shown to form via a reaction of the methoxy radical with \ce{CH2OH}\citep{gar08}. The methoxy radical could feasibly react with other organic radicals to form the methoxylated versions of various additional organic species. Therefore, the methoxylated counterparts of the high abundance organics in this source could be interesting radioastronomical targets. However, the microwave spectrum of methoxyethanol has only been experimentally collected and assigned up to 26.5 GHz \citep{buc72}.

\subsubsection{Ethane (\ce{C2H6}), Carbon Dioxide (\ce{CO2}), and Carbon Disulfide (\ce{CS2})}
For an allowed pure rotational spectrum to be collected, a molecule must have a nonzero dipole moment. Therefore, nonpolar molecules like ethane, carbon dioxide, and carbon disulfide will be undetectable through radioastronomy based on allowed pure rotational transitions regardless of their high predicted column densities. However, ethane was first detected toward the comet C/1996 B2 Hyakutake with high-resolution infrared spectroscopy \citep{mum96}. Both solid and gaseous \ce{CO2} have also been detected toward various interstellar sources using IR techniques\citep{dhe_co2_89, van_co2_96}. \ce{CS2} has not been previously detected in the interstellar medium. That said, experimental studies have shown that \ce{CS2} can react with oxygen atoms on solid surfaces under astrophysically relevant conditions to form carbonyl sulfide (OCS), which is detected in high abundance toward IRAS 16293B \citep{ward_12_cs2,dro18}  

\subsubsection{3-Hydroxypropanal (\ce{OHCH2CH2CHO})}
Hydroxyacetone (\ce{CH3COCH2OH}), a structural isomer of 3-hydroxypropanal, was detected with a column density of \num{1.2e16} \si{\per\centi\meter\squared} toward IRAS 16293B \citep{zho20}. Experiments by \citet{wan23} showed that 3-hydroxypropanal can be formed in methanol-acetaldehyde ices irradiated with energetic electrons at 5\,K. They concluded that this molecule can be produced in interstellar ices of star-forming regions that have high abundances of methanol and acetaldehyde (which is the case in IRAS 16293B). This molecule would then be desorbed into the gas phase as the protostar is heated. However, to our knowledge the rotational spectrum of 3-hydroxypropanal has not been experimentally measured.

\subsection{Isotope Ratios}
\label{sec:isotopes}

Since isotopic composition was encoded in our molecular feature vectors, we proceeded to test the regressors' ability to predict isotopic ratios. As noted previously, if the machine learning model can accurately predict isotope ratios, information about the evolutionary history of the source and the molecular formation can be deciphered. That said, with such a simple encoding of the isotopic information in our feature vectors as well as the relatively small collection of isotopically substituted molecules, modeling this nuanced chemistry will be a challenge. Our discussion will solely focus on \isotope[13]{C} and D substituted isotopologues because of the extremely small sample size of all other minor isotopes.   

\begin{figure*}[h!]
\includegraphics[width=\textwidth]{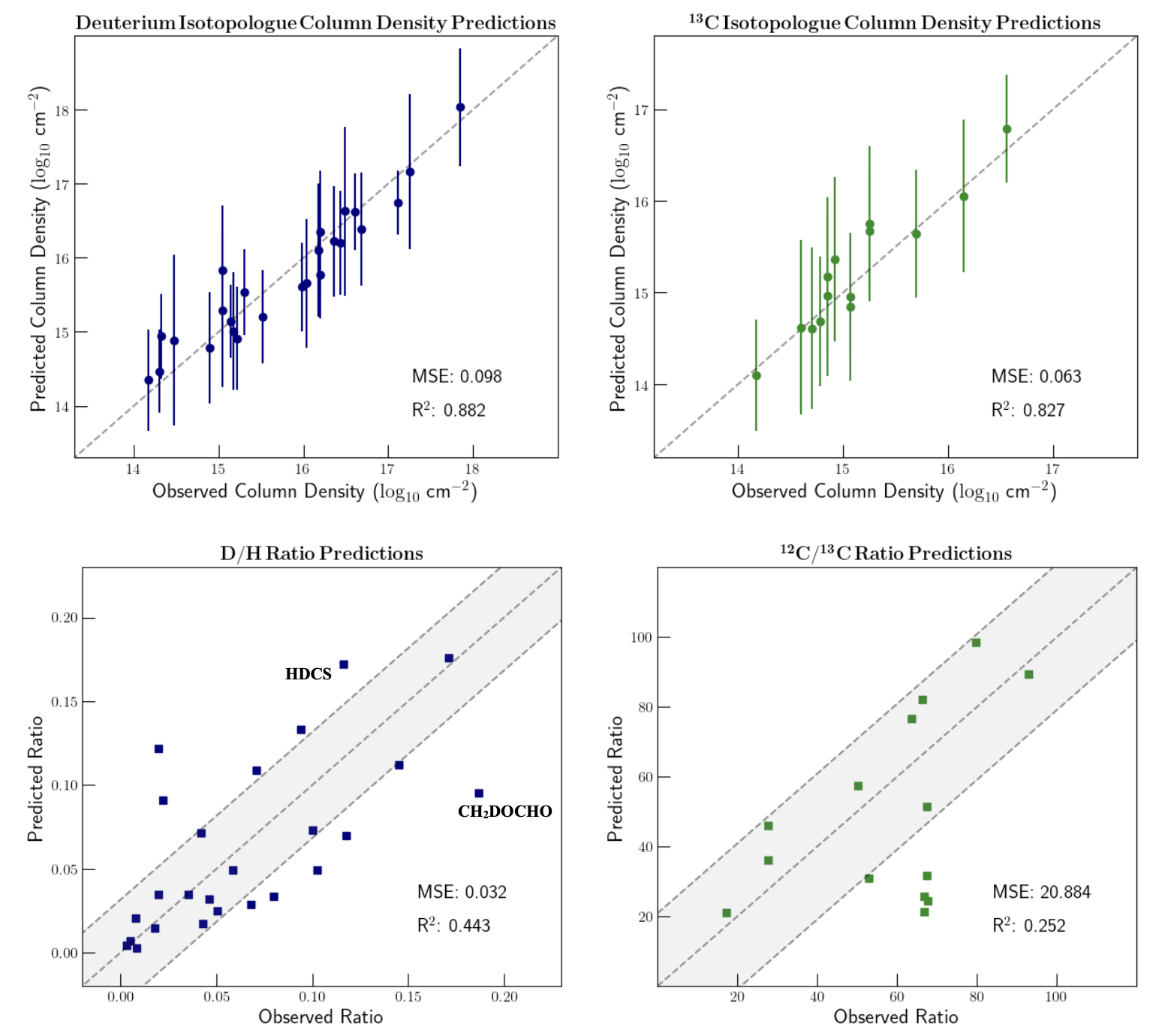}
\caption{GPR column density and isotope ratio predictions of deuterium and \isotope[13]{C} substituted isotopologues using hand-picked isotope features. 1$\sigma$ column density uncertainties are shown. Some of the notable prediction errors are labelled and discussed in the text. Points that are within the grey shading denote molecules for which the ratio prediction error is less than the mean absolute error.  }\label{fig5}
\end{figure*}

Figure \ref{fig5} displays the predicted column densities of the D and \isotope[13]{C} substituted isotopologues along with the corresponding isotopic ratios. These predictions stem from 5-fold cross validation on the isotopically substituted data. In this process, the isotopologues are split into five subsets of training and validation data. In each iteration, 20\% of the isotopically substituted molecules are left out of the training set. The model is then trained on all molecules in the dataset besides the 20\% of rare isotopologues that were assigned to the validation set. 

Because the deuterium and \isotope[13]{C} ratios are reported in inverse fashions, the mean squared errors of the two ratio plots differ dramatically. The points within the shaded regions denote the molecules for which the prediction error is less than the mean absolute prediction error. For the D/H ratios, the mean absolute error is 0.032. For the \isotope[12]{C}/\isotope[13]{C} ratios, the mean absolute error is 20.9.

The column density predictions for isotopically substituted molecules are typically extremely accurate. However, when considering isotopic ratios, the range of realistic values is quite limited; therefore, a small prediction error is very notable. For example, deuterated acetaldehyde in IRAS 16293B is observed to have a D/H ratio of 7.98\%. A column density under-prediction by 0.3 orders of magnitude would result in a predicted ratio of approximately 4.00\%. This ratio suggests very different temperatures and timescales of formation. 

There are a few molecules for which the ratio prediction is especially inaccurate. These species are labelled on Figure \ref{fig5}. The prediction errors of \ce{HDCS} and \ce{CH2DOCHO} are especially notable since they highlight the shortcomings of using hand-picked descriptors. When encoding the chemical environment of the deuterium substitution, vector dimensions are included that denote whether there is a carbon or oxygen atom two bonds away from the deuterium atom. However, there was no consideration of whether the atom in this position is sulfur since \ce{HDCS} is the only molecule in the dataset in which this is the case. Therefore, this important chemical environment information is not provided to the model, thus leading to a large prediction error. Additionally, with simple hand-picked features, the model isn't always able to capture the nuances of isotopic fractionation. For example, the isotopic encoding of \ce{CH2DOCHO} is very similar to that of \ce{CH2DOH}. Because \ce{CH2DOH} has a D/H ratio of only around 7\%, the model inaccurately predicts \ce{CH2DOCHO} to have approximately the same ratio.

\begin{figure*}[h!]
\includegraphics[width=\textwidth]{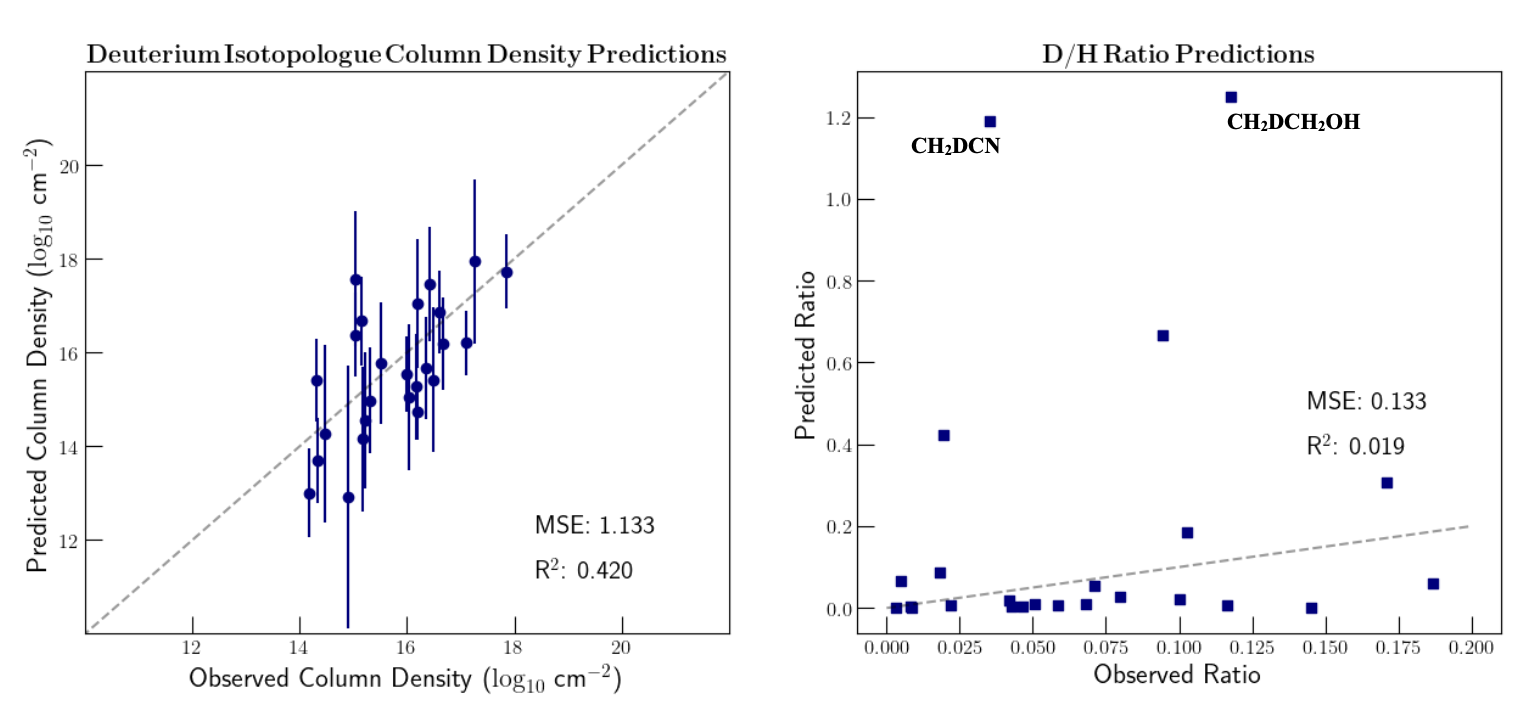}
\caption{GPR column density and isotope ratio predictions of deuterium substituted isotopologues. The inputted molecular feature vectors were generated with the \textsc{Mol2vec} algorithm and had no additional isotopic information included. 1$\sigma$ column density uncertainties are shown. Some of the notable prediction errors are labelled and discussed in the text.}\label{fig6}
\end{figure*}

Preferably we could include a more nuanced encoding of isotopic composition that better captures the local chemical environment instead of simple hand-picked features. However, with only 27 deuterated molecules and 15 \isotope[13]{C} substituted species, the dataset of unique isotopologues is too small to learn the required relationships with a complex featurization. As mentioned previously, \textsc{Mol2vec} is sensitive to some, but not all, isotopic substitutions. It can, however, create unique vectors for deuterated species. Therefore, we tested the ability to learn deuterium ratios from the original \textsc{Mol2vec}-produced vectors that more fully consider chemical context. These results are shown in Figure \ref{fig6}. The D/H ratio of formic acid is omitted from the graph since a ratio of around 6 is predicted which skews the ability to view the remaining points. As can be seen, these predictions are far less accurate than when hand-picked features were used.  This is because the vector representations of many of the deuterated species are quite dissimilar to the main isotopologue in this case. In fact, the vector representation of \ce{CH2DCH2OH} is closer to that of propanal than that of \ce{CH3CH2OH}. In order to include more detailed isotopic information in the feature vectors and thus accurately model isotopic fractionation, it is clear that we require additional isotopologue detections.

Finally, in order to test the predictive ability of the model on isotopologues for which no column density has been derived, we proceeded to predict the  \isotope[12]{C}/\isotope[13]{C} ratio of CO and CN with the trained GPR model from Section \ref{sec:regression}. For these two molecules (along with \ce{H2CO}), a linear \isotope[12]{C}/\isotope[13]{C} trend has been defined as a function of Galactocentric distance (D\textsubscript{GC}). The formulae of these Galactocentric gradients are displayed in Equations \ref{form:co}, \ref{form:cn}, and \ref{form:formaldehyde}

\begin {equation}
  \ce{^{12}CO}/\ce{^{13}CO} = (5.41 \pm 1.07) \textrm{kpc\textsuperscript{-1}} \times  \textrm{D\textsubscript{GC}} + (19.03 \pm 7.90)
  \label{form:co}
\end {equation}

\begin {equation}
  \ce{^{12}CN}/\ce{^{13}CN} = (6.01\pm 1.19) \textrm{kpc\textsuperscript{-1}} \times  \textrm{D\textsubscript{GC}} + (12.28 \pm 9.33)
  \label{form:cn}
\end {equation}

\begin {equation}
  \ce{H2}\ce{^{12}CO}/\ce{H2}\ce{^{13}CO} = (7.60 \pm 1.79) \textrm{kpc\textsuperscript{-1}} \times  \textrm{D\textsubscript{GC}} + (18.05 \pm 10.88)
  \label{form:formaldehyde}
\end {equation}

Using a Galactocentric distance of 8.043 kpc for IRAS 16293, the range of expected ratios along with the ratios predicted by the GPR model are shown in Table \ref{isotopeRatios}. For reference, the observed \isotope[12]{C}/\isotope[13]{C} ratio of \ce{H2CO} is listed as well \citep{per18}. The Galactocentric distance of IRAS 16293 was computed using the \textsc{astropy} Python module \citep{astropy}. The values used in this calculation were the distances from the Earth to both the Galactic Center and IRAS 16293 (8.178 kpc and 141 pc\citep{dgc, iras_distance}) as well as their respective sky coordinates.

\begin{table*}[h!]

\begin{center}
\caption{\label{isotopeRatios} Measured and predicted \isotope[12]{C}/\isotope[13]{C} ratios of CO, CN, and \ce{H2CO} toward IRAS 16293B along with the expected ratios that correspond to the \isotope[13]{C} Galactocentric gradient.}
\begin{tabular}{cccc} 
\toprule

Molecule & Galactocentric Gradient \isotope[12]{C}/\isotope[13]{C} Ratio  &  Predicted  \isotope[12]{C}/\isotope[13]{C} Ratio & Observed \isotope[12]{C}/\isotope[13]{C} \\ [1ex] 
 \midrule

CO &  46.04 -- 79.05 & 50.85 & -- \\
CN &  41.72 -- 79.52 & 27.34 & -- \\
\ce{H2CO} &  53.90 -- 104.46 & -- & 52.92 \\
 \bottomrule

\end{tabular}
\end{center}
\end{table*}

The observed \isotope[12]{C}/\isotope[13]{C} ratio of formaldehyde toward IRAS 16293B is very near the lower bound of the Galactocentric trend error bars at 8.043 kpc. Interestingly, the GPR model predicts the CO and CN ratios to also be fairly near the lower bounds of the respective ratio gradients. This matches the observed trends of various other molecules in IRAS 16293B, which typically show high levels of \isotope[13]{C} substitution. A high abundance of \isotope[13]{C}O could stem from favorable ion-neutral isotope exchange reactions in the cold interstellar gas before CO freeze-out \citep{lan84}. Complex organic species that were then formed on grain surfaces from CO following freeze-out would inherit this small \isotope[12]{C}/\isotope[13]{C} ratio. The \isotope[13]{C} enhancement in organic molecules would be even more notable at later timescales since laboratory experiments have shown that \ce{^{12}CO} desorbs slightly more efficiently than \ce{^{13}CO} \citep{smi21}. This enables \ce{^{12}CO} to sublimate from the grain at a lower temperature. Therefore, as the protostar begins to heat the surroundings to around the CO sublimation temperature, the more efficient \ce{^{12}CO} sublimation would result in grain surfaces that are further enhanced with \ce{^{13}CO}. 

Overall, while the machine learning regressor is not precise enough to adequately model the exact isotopic ratios, the \isotope[13]{C}O and \isotope[13]{C}N predictions show that it still is able to learn the general overabundance of \isotope[13]{C} in the organic species of IRAS 16293B.

\subsection{Isotopologue targets for astrochemical study}
\label{sec:iso_predict}

\begin{figure}[h]
\includegraphics[width=\columnwidth]{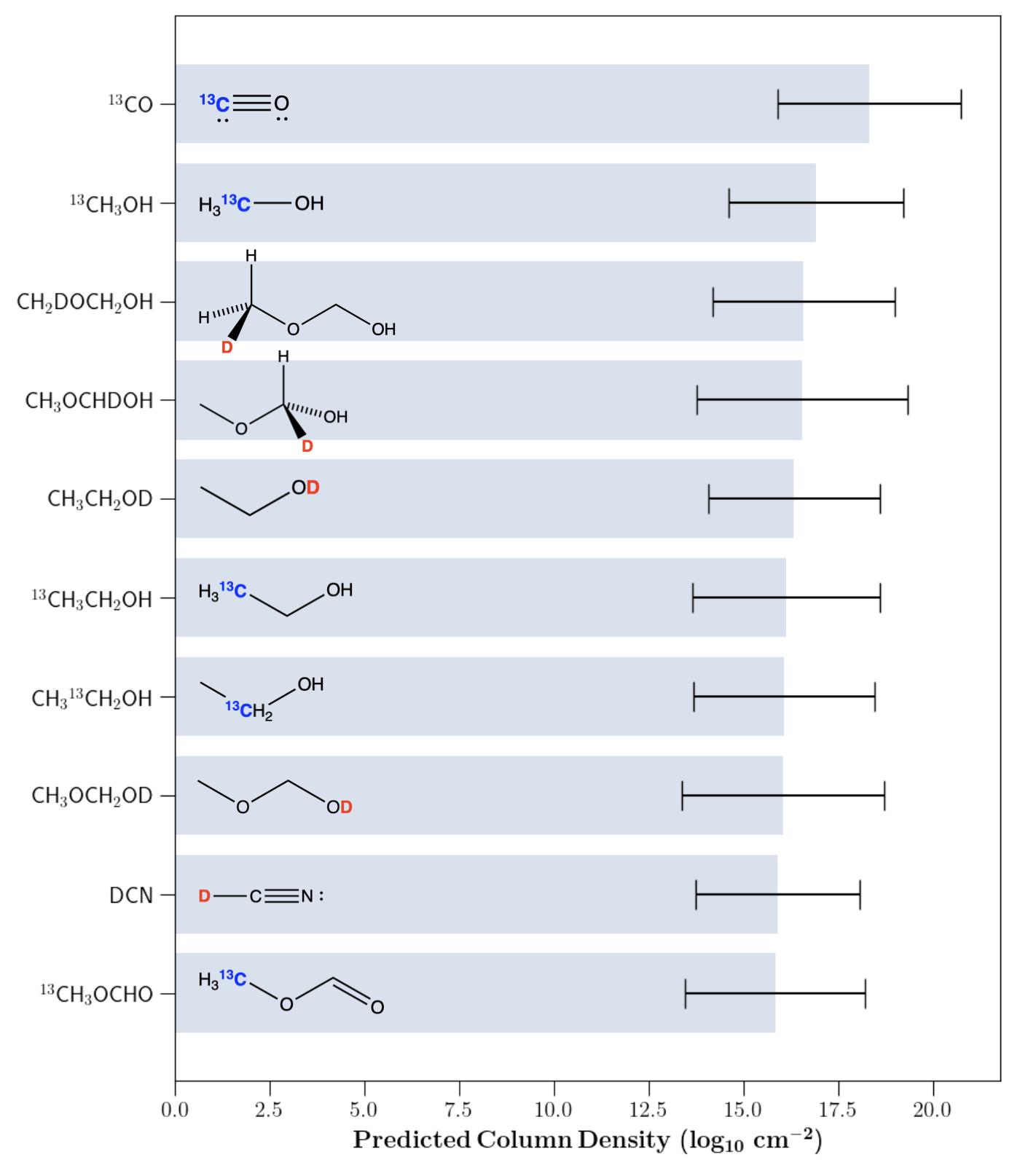}
\caption{The 10 undetected rare isotopologues of the molecules identified in IRAS 16293B with the highest predicted column densities. The trained Gaussian process regression model was used for the predictions. 1$\sigma$ column density uncertainties are shown. }\label{isotope_predictions}
\end{figure}

As noted in Section \ref{sec:intro}, because of the extreme isotopic fractionation in this source, multiple groups have predicted that a significant portion of the unidentified spectral peaks in the line survey arise from isotopically substituted species \citep{jor16, taq18_o2}. Additionally,  Figure \ref{fig5} shows that the GPR regressor is able to precisely model column densities of the isotopically substituted molecules. Therefore, using the trained GPR model we proceeded to predict the column densities of the D and \isotope[13]{C} mono-substituted isotopologues of the previously identified molecules in IRAS 16293B for which no accurate column density has been derived. Due to the inaccurate predictions of the main isotopologues of ethylene glycol, chloromethane, and dimethyl ether, the predictions of the rare isotopologues of these species were omitted. The highest 10 predicted column density isotopologues are displayed in Figure \ref{isotope_predictions}. All predictions are provided in the associated \href{https://github.com/zfried/IRAS_ML_Predictions}{GitHub repository}.  

Of these highest 10 predicted column density species, both \isotope[13]{C}-substituted ethanol isotopologues and OD substituted ethanol were marginally detected toward IRAS 16293B. While we did not include these molecules in our training set, \citet{jor18} derived tentative column densities for these species. These tentatively derived column densities along with the values predicted by the GPR model are listed in Table \ref{tentative}. As can be seen, all predictions closely match the tentative column densities. Additionally, while transitions corresponding to \isotope[13]{C}\ce{H3OH} and DCN have been identified toward IRAS 16293, no derived column density is listed \citep{jor11_iso}.

\begin{table*}[h!]
\begin{center}
\caption{\label{tentative} Tentative column densities of various ethanol isotopologues derived by \citet{jor18} corresponding to marginal detections along with the column densities predicted by the trained Gaussian process regression model.}
\begin{tabular}{ccc} 
\toprule

Molecule & Tentative Column Density (\si{\per\centi\meter\squared})  &  Predicted Column Density (\si{\per\centi\meter\squared}) \\ [1ex] 
 \midrule
 \ce{CH3CH2OD} & \num{1.1e16} & \num{2.15e16} \\
\isotope[13]{C}\ce{H3CH2OH} & \num{9.1e15}  & \num{1.33e16}  \\
\ce{CH3}\isotope[13]{C}\ce{H2OH} &  \num{9.1e15} & \num{1.18e16}  \\

 \bottomrule

\end{tabular}
\end{center}
\end{table*}

Therefore, the remaining molecules of interest are the three deuterated isotopologues of methoxymethanol along with the \isotope[13]{C} isotopologue of methyl formate. Beyond these largest 10 predicted column density isotopologues, another high predicted abundance isotopologue is the deuterated isotopologue of methoxyethane (\ce{CH3CH2OCH2D}). As mentioned previously, while large column densities are beneficial for detection, various other factors impact the detectability of a molecule. For all of these molecules, we  therefore simulated their spectra toward IRAS 16293B using the predicted column densities in order to assess their true detectability. The microwave and sub-mm spectra of \isotope[13]{C}\ce{H3OCHO} have been experimentally studied and assigned, thus making interstellar detection currently possible. In fact, this rare isotopologue was detected in the Orion molecular cloud \citep{formate_iso}. However, this particular isotopologue is not present in the CDMS molecular spectroscopy database \citep{cdms}. All of the other aforementioned deuterated isotopologues have not been studied experimentally. 

In order to simulate the spectra of these previously unstudied isotopologues, the rotational constants must first be calculated. A low-cost method to obtain molecular rotational constants for isotopically substituted species can be achieved by combining experimental data and ab initio calculations. In this process, the experimental rotational and distortion constants of the main isotopologue are first collected. The $A$, $B$, and $C$ rotational constants of the parent species are then calculated at a given level of theory and basis set. For our work, we ran the calculations with the PSI4 \citep{psi4} Python package and used the M06-2X functional with the 6-311++G(d,p) basis set. Assuming that the geometry remains constant upon isotopic substitution, the same computational methods are then used to calculate rotational constants of the rare isotopologues. Finally, it is assumed that the scaling factor between the experimental and calculated rotational constants is the same for the main isotopologues and the isotopically substituted molecules. For example, the $B$ rotational constant of the isotopically substituted species can be calculated using Equation \ref{form:scaling}.

\begin {equation}
  B\textrm{\textsubscript{scaled}} = \frac{B\textrm{\textsubscript{exp} (parent)}} {B\textrm{\textsubscript{calc} (parent)}} \times  B\textrm{\textsubscript{calc}} 
  \label{form:scaling}
\end {equation}

The experimental distortion constants and dipole moments of the main isotopologues were used as-is for the isotopically substituted molecules.

Following the rotational constant calculations, a rotational line catalog was generated using Pickett's SPCAT \citep{spcat}. Only the $A$, $B$, and $C$ rotational constants and distortion constants (when available) were used. Internal rotational was not considered during the  catalog simulations. The \texttt{molsim} \citep{molsim} Python package was then used to simulate the spectra of the isotopologues toward IRAS 16293B with the predicted column densities. \texttt{molsim} assumes that the molecular emission can be described by a single excitation temperature, and accounts for the effects of optical depth. For the simulations, the excitation temperature and v\textsubscript{lsr} of the main isotopologue were used. A source size of 0.5$^{\prime\prime}$, beam diameter of 0.5$^{\prime\prime}$, and line width of 1.0 km/s were used for each simulation.

\begin{figure*}
\begin{center}
\includegraphics[width=\textwidth]{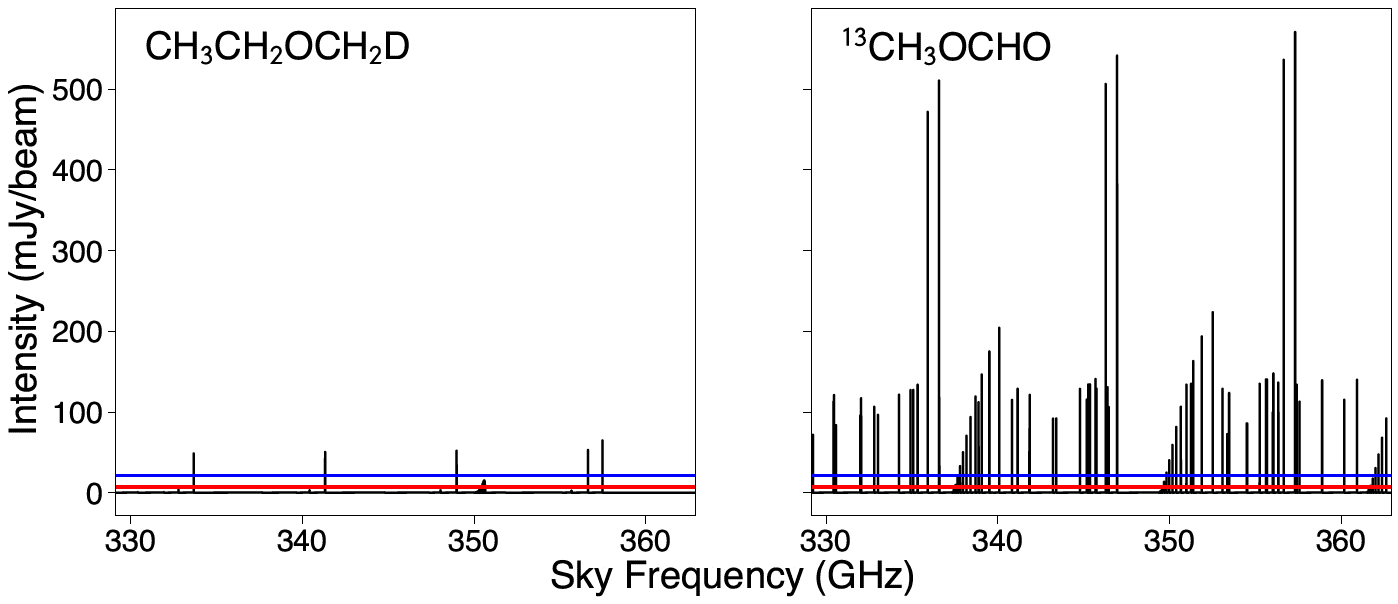}
\caption{Simulated spectra of \ce{CH3CH2OCH2D} and \isotope[13]{C}\ce{H3OCHO} toward IRAS 16293B. The column densities used for the simulations were \num{3.98e15} \si{\per\centi\meter\squared} for \ce{CH3CH2OCH2D} and \num{6.80e15} \si{\per\centi\meter\squared} for \isotope[13]{C}\ce{H3OCHO}. These values were predicted by the trained Gaussian process regression model. The red line denotes the approximate RMS noise level of the PILS observations in ALMA Band 7. The blue line denotes the intensity required for a transition to have 3$\sigma$ significance. Other simulation parameters are noted in the text. }\label{simspectra}
\end{center}
\end{figure*}

The frequency range of the PILS observations in ALMA Band 7 (329.147-362.896 GHz) was the predominant focus of our spectral simulations. Given the noise level of the PILS observations \citep{jor16}, any transition with a peak intensity stronger than $\sim$21 mJy/beam should be detectable at a 3$\sigma$ significance. Analysis of the spectral simulations can be seen in Table \ref{isotopologue_predictions}. The simulated spectra of the \isotope[13]{C}-substituted methyl formate isotopologue and the deuterated methoxyethane isotopologue toward IRAS 16293B are presented in Figure \ref{simspectra}. 

\begin{table*}[h!]

\begin{center}
\caption{\label{isotopologue_predictions} Analysis of the simulated spectra of the three deuterated isotopologues of methoxymethanol, \isotope[13]{C} substituted methyl formate, and deuterated methoxyethane. The spectral line catalogs were produced using SPCAT and the spectra were simulated with the \texttt{molsim} Python module. The simulations span part of ALMA Band 7 (329.147 - 362.896 GHz). 1$\sigma$ transitions have intensities of 7 mJy/beam and 3$\sigma$ transitions have intensities of 21 mJy/beam.}
\begin{tabular}{cccc} 
\toprule

Molecule & \# of 3$\sigma$ Transitions & \# of 1$\sigma$ Transitions & Intensity of Strongest Transition (mJy/beam) \\ [1ex] 
 \midrule
\isotope[13]{C}\ce{H3OCHO} & 81  & 86 & 571.0  \\
\ce{CH3CH2OCH2D} &  12 & 18 & 64.8  \\
\ce{CH2DOCH2OH} &  0 & 0 & 0.39 \\
\ce{CH3OCHDOH} &  0 & 0 & 0.18 \\
\ce{CH3OCH2OD} & 0 & 0 & \num{3.92e-8} \\

 \bottomrule

\end{tabular}
\end{center}
\end{table*}

Despite having higher predicted column densities, the rare isotopologues of methoxymethanol are predicted to have much weaker spectral peaks than the other isotopologues considered due to the limited dipole moment. That said, with several transitions predicted to be stronger than 3$\sigma$, \ce{CH3CH2OCH2D} is an excellent candidate for experimental study for astrochemical purposes. Additionally, since the spectrum of \isotope[13]{C}\ce{H3OCHO} has already been collected and assigned, we recommend that this molecule be searched for in the PILS data.  

\section{Conclusions}

In this work, we applied the machine learning pipeline introduced by \citet{lee21} to source B of the Class 0 protostellar system IRAS 16293-2422. In order to include isotopologues in the dataset, we also concatenated a simple encoding of the isotopic composition to the feature vectors. Gaussian process regression and Bayesian ridge regression were both able to accurately model the column densities of the detected molecules in this source. The trained Gaussian process regression model then provided a list of 242 well-constrained targets for astrochemical study. Small, oxygenated, and fairly saturated hydrocarbons were predicted to be in high abundance in this protostellar source. While the column density predictions of isotopologues were quite precise, the nuances of isotopic ratios were only modeled with moderate accuracy. Additional isotopologue detections will be required to allow for a more complex encoding of isotopic substitution that better captures local chemical environment. Finally, since it has been predicted that many of the unassigned transitions in the PILS survey arise from isotopically substituted molecules, we provided a list of 92 isotopologue column density predictions. 

This machine learning method has now been shown to effectively model the molecular column densities in two separate interstellar sources and the resulting trained regression models can be used to predict molecular species that are likely abundant in these various regions of interstellar space. However, these same techniques can be readily applied to terrestrial chemical mixture identifications as well. For example, if a researcher is able to reliably identify a fairly small number of chemical components present in an environmental sample along with their abundances, these supervised regressors could be trained and used to predict other components and contaminants of that mixture.  

\section*{Author Contributions}
\textbf{Zachary Fried: }Software, Formal analysis, Investigation, Data Curation, Writing - Original Draft, Visualization. \textbf{Kelvin Lee: }Methodology, Supervision, Writing - Review \& Editing. \textbf{Alex Byrne: }Software, Writing - Review \& Editing.  \textbf{Brett McGuire: }Supervision, Writing - Review \& Editing, Funding acquisition.

\section*{Conflicts of interest}
There are no conflicts to declare.

\section*{Acknowledgements}
The authors thank C. W. Coley for providing helpful insight regarding the \textsc{Mol2vec} molecular featurization and the inclusion of isotopic information in Morgan fingerprints.

The National Radio Astronomy Observatory is a facility of the National Science Foundation operated under cooperative agreement by Associated Universities, Inc.

\bibliographystyle{rsc}
\bibliography{references}

\providecommand*{\mcitethebibliography}{\thebibliography}
\csname @ifundefined\endcsname{endmcitethebibliography}
{\let\endmcitethebibliography\endthebibliography}{}
\begin{mcitethebibliography}{84}
\providecommand*{\natexlab}[1]{#1}
\providecommand*{\mciteSetBstSublistMode}[1]{}
\providecommand*{\mciteSetBstMaxWidthForm}[2]{}
\providecommand*{\mciteBstWouldAddEndPuncttrue}
  {\def\EndOfBibitem{\unskip.}}
\providecommand*{\mciteBstWouldAddEndPunctfalse}
  {\let\EndOfBibitem\relax}
\providecommand*{\mciteSetBstMidEndSepPunct}[3]{}
\providecommand*{\mciteSetBstSublistLabelBeginEnd}[3]{}
\providecommand*{\EndOfBibitem}{}
\mciteSetBstSublistMode{f}
\mciteSetBstMaxWidthForm{subitem}
{(\emph{\alph{mcitesubitemcount}})}
\mciteSetBstSublistLabelBeginEnd{\mcitemaxwidthsubitemform\space}
{\relax}{\relax}

\bibitem[van Dishoeck and Blake(1998)]{van98}
E.~F. van Dishoeck and G.~A. Blake, \emph{Annual Review of Astronomy and
  Astrophysics}, 1998, \textbf{36}, 317--368\relax
\mciteBstWouldAddEndPuncttrue
\mciteSetBstMidEndSepPunct{\mcitedefaultmidpunct}
{\mcitedefaultendpunct}{\mcitedefaultseppunct}\relax
\EndOfBibitem
\bibitem[Öberg \emph{et~al.}(2011)Öberg, Murray-Clay, and Bergin]{obe11}
K.~I. Öberg, R.~Murray-Clay and E.~A. Bergin, \emph{The Astrophysical
  Journal}, 2011, \textbf{743}, L16\relax
\mciteBstWouldAddEndPuncttrue
\mciteSetBstMidEndSepPunct{\mcitedefaultmidpunct}
{\mcitedefaultendpunct}{\mcitedefaultseppunct}\relax
\EndOfBibitem
\bibitem[Bachiller \emph{et~al.}(2001)Bachiller, Pérez~Gutiérrez, Kumar, and
  Tafalla]{bac01}
R.~Bachiller, M.~Pérez~Gutiérrez, M.~S.~N. Kumar and M.~Tafalla,
  \emph{Astronomy \& Astrophysics}, 2001, \textbf{372}, 899--912\relax
\mciteBstWouldAddEndPuncttrue
\mciteSetBstMidEndSepPunct{\mcitedefaultmidpunct}
{\mcitedefaultendpunct}{\mcitedefaultseppunct}\relax
\EndOfBibitem
\bibitem[Schilke \emph{et~al.}(1997)Schilke, Walmsley, Pineau~des Forets, and
  Flower]{sch97}
P.~Schilke, C.~M. Walmsley, G.~Pineau~des Forets and D.~R. Flower,
  \emph{Astronomy \& Astrophysics}, 1997, \textbf{321}, 293--304\relax
\mciteBstWouldAddEndPuncttrue
\mciteSetBstMidEndSepPunct{\mcitedefaultmidpunct}
{\mcitedefaultendpunct}{\mcitedefaultseppunct}\relax
\EndOfBibitem
\bibitem[Ginsburg \emph{et~al.}(2019)Ginsburg, McGuire, Plambeck, Bally, Goddi,
  and Wright]{gin19}
A.~Ginsburg, B.~McGuire, R.~Plambeck, J.~Bally, C.~Goddi and M.~Wright,
  \emph{The Astrophysical Journal}, 2019, \textbf{872}, 54\relax
\mciteBstWouldAddEndPuncttrue
\mciteSetBstMidEndSepPunct{\mcitedefaultmidpunct}
{\mcitedefaultendpunct}{\mcitedefaultseppunct}\relax
\EndOfBibitem
\bibitem[Ho \emph{et~al.}(1979)Ho, Barrett, Myers, Matsakis, Cheung, Chui,
  Townes, and Yngvesson]{ho79}
P.~T.~P. Ho, A.~H. Barrett, P.~C. Myers, D.~N. Matsakis, A.~L. Cheung, M.~F.
  Chui, C.~H. Townes and K.~S. Yngvesson, \emph{The Astrophysical Journal},
  1979, \textbf{234}, 912--921\relax
\mciteBstWouldAddEndPuncttrue
\mciteSetBstMidEndSepPunct{\mcitedefaultmidpunct}
{\mcitedefaultendpunct}{\mcitedefaultseppunct}\relax
\EndOfBibitem
\bibitem[Ruaud \emph{et~al.}(2016)Ruaud, Wakelam, and Hersant]{rua16}
M.~Ruaud, V.~Wakelam and F.~Hersant, \emph{Monthly Notices of the Royal
  Astronomical Society}, 2016, \textbf{459}, 3756--3767\relax
\mciteBstWouldAddEndPuncttrue
\mciteSetBstMidEndSepPunct{\mcitedefaultmidpunct}
{\mcitedefaultendpunct}{\mcitedefaultseppunct}\relax
\EndOfBibitem
\bibitem[Wakelam \emph{et~al.}(2015)Wakelam, Loison, Herbst, Pavone, Bergeat,
  Béroff, Chabot, Faure, Galli, Geppert, Gerlich, Gratier, Harada, Hickson,
  Honvault, Klippenstein, Picard, Nyman, Ruaud, Schlemmer, Sims, Talbi,
  Tennyson, and Wester]{wak15}
V.~Wakelam, J.-C. Loison, E.~Herbst, B.~Pavone, A.~Bergeat, K.~Béroff,
  M.~Chabot, A.~Faure, D.~Galli, W.~D. Geppert, D.~Gerlich, P.~Gratier,
  N.~Harada, K.~M. Hickson, P.~Honvault, S.~J. Klippenstein, S.~D.~L. Picard,
  G.~Nyman, M.~Ruaud, S.~Schlemmer, I.~R. Sims, D.~Talbi, J.~Tennyson and
  R.~Wester, \emph{The Astrophysical Journal Supplement Series}, 2015,
  \textbf{217}, 20\relax
\mciteBstWouldAddEndPuncttrue
\mciteSetBstMidEndSepPunct{\mcitedefaultmidpunct}
{\mcitedefaultendpunct}{\mcitedefaultseppunct}\relax
\EndOfBibitem
\bibitem[McGuire \emph{et~al.}(2015)McGuire, Carroll, Dollhopf, Crockett,
  Corby, Loomis, Burkhardt, Shingledecker, Blake, and Remijan]{mcg15}
B.~A. McGuire, P.~B. Carroll, N.~M. Dollhopf, N.~R. Crockett, J.~F. Corby,
  R.~A. Loomis, A.~M. Burkhardt, C.~Shingledecker, G.~A. Blake and A.~J.
  Remijan, \emph{The Astrophysical Journal}, 2015, \textbf{812}, 76\relax
\mciteBstWouldAddEndPuncttrue
\mciteSetBstMidEndSepPunct{\mcitedefaultmidpunct}
{\mcitedefaultendpunct}{\mcitedefaultseppunct}\relax
\EndOfBibitem
\bibitem[Hébrard \emph{et~al.}(2009)Hébrard, Dobrijevic, Pernot, Carrasco,
  Bergeat, Hickson, Canosa, Le~Picard, and Sims]{heb09}
E.~Hébrard, M.~Dobrijevic, P.~Pernot, N.~Carrasco, A.~Bergeat, K.~M. Hickson,
  A.~Canosa, S.~D. Le~Picard and I.~R. Sims, \emph{The Journal of Physical
  Chemistry A}, 2009, \textbf{113}, 11227--11237\relax
\mciteBstWouldAddEndPuncttrue
\mciteSetBstMidEndSepPunct{\mcitedefaultmidpunct}
{\mcitedefaultendpunct}{\mcitedefaultseppunct}\relax
\EndOfBibitem
\bibitem[Lee \emph{et~al.}(2021)Lee, Patterson, Burkhardt, Vankayalapati,
  McCarthy, and McGuire]{lee21}
K.~L.~K. Lee, J.~Patterson, A.~M. Burkhardt, V.~Vankayalapati, M.~C. McCarthy
  and B.~A. McGuire, \emph{The Astrophysical Journal Letters}, 2021,
  \textbf{917}, L6\relax
\mciteBstWouldAddEndPuncttrue
\mciteSetBstMidEndSepPunct{\mcitedefaultmidpunct}
{\mcitedefaultendpunct}{\mcitedefaultseppunct}\relax
\EndOfBibitem
\bibitem[Wootten(1989)]{woo89}
A.~Wootten, \emph{The Astrophysical Journal}, 1989, \textbf{337}, 858\relax
\mciteBstWouldAddEndPuncttrue
\mciteSetBstMidEndSepPunct{\mcitedefaultmidpunct}
{\mcitedefaultendpunct}{\mcitedefaultseppunct}\relax
\EndOfBibitem
\bibitem[Mundy \emph{et~al.}(1992)Mundy, Wootten, Wilking, Blake, and
  Sargent]{mun92}
L.~G. Mundy, A.~Wootten, B.~A. Wilking, G.~A. Blake and A.~I. Sargent,
  \emph{The Astrophysical Journal}, 1992, \textbf{385}, 306\relax
\mciteBstWouldAddEndPuncttrue
\mciteSetBstMidEndSepPunct{\mcitedefaultmidpunct}
{\mcitedefaultendpunct}{\mcitedefaultseppunct}\relax
\EndOfBibitem
\bibitem[Looney \emph{et~al.}(2000)Looney, Mundy, and Welch]{loo00}
L.~W. Looney, L.~G. Mundy and W.~J. Welch, \emph{The Astrophysical Journal},
  2000, \textbf{529}, 477--498\relax
\mciteBstWouldAddEndPuncttrue
\mciteSetBstMidEndSepPunct{\mcitedefaultmidpunct}
{\mcitedefaultendpunct}{\mcitedefaultseppunct}\relax
\EndOfBibitem
\bibitem[Maureira \emph{et~al.}(2020)Maureira, Pineda, Segura-Cox, Caselli,
  Testi, Lodato, Loinard, and Hernández-Gómez]{mau20}
M.~J. Maureira, J.~E. Pineda, D.~M. Segura-Cox, P.~Caselli, L.~Testi,
  G.~Lodato, L.~Loinard and A.~Hernández-Gómez, \emph{The Astrophysical
  Journal}, 2020, \textbf{897}, 59\relax
\mciteBstWouldAddEndPuncttrue
\mciteSetBstMidEndSepPunct{\mcitedefaultmidpunct}
{\mcitedefaultendpunct}{\mcitedefaultseppunct}\relax
\EndOfBibitem
\bibitem[Jørgensen \emph{et~al.}(2016)Jørgensen, Wiel, Coutens, Lykke,
  Müller, Dishoeck, Calcutt, Bjerkeli, Bourke, Drozdovskaya, Favre, Fayolle,
  Garrod, Jacobsen, Öberg, Persson, and Wampfler]{jor16}
J.~K. Jørgensen, M.~H. D. v.~d. Wiel, A.~Coutens, J.~M. Lykke, H.~S.~P.
  Müller, E.~F.~v. Dishoeck, H.~Calcutt, P.~Bjerkeli, T.~L. Bourke, M.~N.
  Drozdovskaya, C.~Favre, E.~C. Fayolle, R.~T. Garrod, S.~K. Jacobsen, K.~I.
  Öberg, M.~V. Persson and S.~F. Wampfler, \emph{Astronomy \& Astrophysics},
  2016, \textbf{595}, A117\relax
\mciteBstWouldAddEndPuncttrue
\mciteSetBstMidEndSepPunct{\mcitedefaultmidpunct}
{\mcitedefaultendpunct}{\mcitedefaultseppunct}\relax
\EndOfBibitem
\bibitem[Taquet \emph{et~al.}(2018)Taquet, Dishoeck, Swayne, Harsono,
  Jørgensen, Maud, Ligterink, Müller, Codella, Altwegg, Bieler, Coutens,
  Drozdovskaya, Furuya, Persson, Hoff, Walsh, and Wampfler]{taq18_o2}
V.~Taquet, E.~F.~v. Dishoeck, M.~Swayne, D.~Harsono, J.~K. Jørgensen, L.~Maud,
  N.~F.~W. Ligterink, H.~S.~P. Müller, C.~Codella, K.~Altwegg, A.~Bieler,
  A.~Coutens, M.~N. Drozdovskaya, K.~Furuya, M.~V. Persson, M.~L. R.~v. Hoff,
  C.~Walsh and S.~F. Wampfler, \emph{Astronomy \& Astrophysics}, 2018,
  \textbf{618}, A11\relax
\mciteBstWouldAddEndPuncttrue
\mciteSetBstMidEndSepPunct{\mcitedefaultmidpunct}
{\mcitedefaultendpunct}{\mcitedefaultseppunct}\relax
\EndOfBibitem
\bibitem[Burkhardt \emph{et~al.}(2018)Burkhardt, Herbst, Kalenskii, McCarthy,
  Remijan, and McGuire]{bur_iso}
A.~M. Burkhardt, E.~Herbst, S.~V. Kalenskii, M.~C. McCarthy, A.~J. Remijan and
  B.~A. McGuire, \emph{Monthly Notices of the Royal Astronomical Society},
  2018, \textbf{474}, 5068--5075\relax
\mciteBstWouldAddEndPuncttrue
\mciteSetBstMidEndSepPunct{\mcitedefaultmidpunct}
{\mcitedefaultendpunct}{\mcitedefaultseppunct}\relax
\EndOfBibitem
\bibitem[Jørgensen \emph{et~al.}(2018)Jørgensen, Müller, Calcutt, Coutens,
  Drozdovskaya, Öberg, Persson, Taquet, Dishoeck, and Wampfler]{jor18}
J.~K. Jørgensen, H.~S.~P. Müller, H.~Calcutt, A.~Coutens, M.~N. Drozdovskaya,
  K.~I. Öberg, M.~V. Persson, V.~Taquet, E.~F.~v. Dishoeck and S.~F. Wampfler,
  \emph{Astronomy \& Astrophysics}, 2018, \textbf{620}, A170\relax
\mciteBstWouldAddEndPuncttrue
\mciteSetBstMidEndSepPunct{\mcitedefaultmidpunct}
{\mcitedefaultendpunct}{\mcitedefaultseppunct}\relax
\EndOfBibitem
\bibitem[Coutens \emph{et~al.}(2016)Coutens, Jørgensen, Wiel, Müller, Lykke,
  Bjerkeli, Bourke, Calcutt, Drozdovskaya, Favre, Fayolle, Garrod, Jacobsen,
  Ligterink, Öberg, Persson, Dishoeck, and Wampfler]{cou16}
A.~Coutens, J.~K. Jørgensen, M.~H. D. v.~d. Wiel, H.~S.~P. Müller, J.~M.
  Lykke, P.~Bjerkeli, T.~L. Bourke, H.~Calcutt, M.~N. Drozdovskaya, C.~Favre,
  E.~C. Fayolle, R.~T. Garrod, S.~K. Jacobsen, N.~F.~W. Ligterink, K.~I.
  Öberg, M.~V. Persson, E.~F.~v. Dishoeck and S.~F. Wampfler, \emph{Astronomy
  \& Astrophysics}, 2016, \textbf{590}, L6\relax
\mciteBstWouldAddEndPuncttrue
\mciteSetBstMidEndSepPunct{\mcitedefaultmidpunct}
{\mcitedefaultendpunct}{\mcitedefaultseppunct}\relax
\EndOfBibitem
\bibitem[Ilyushin \emph{et~al.}(2022)Ilyushin, Müller, Jørgensen, Bauerecker,
  Maul, Bakhmat, Alekseev, Dorovskaya, Vlasenko, Lewen, Schlemmer, Berezkin,
  and Lees]{ily22}
V.~V. Ilyushin, H.~S.~P. Müller, J.~K. Jørgensen, S.~Bauerecker, C.~Maul,
  Y.~Bakhmat, E.~A. Alekseev, O.~Dorovskaya, S.~Vlasenko, F.~Lewen,
  S.~Schlemmer, K.~Berezkin and R.~M. Lees, \emph{Astronomy \& Astrophysics},
  2022, \textbf{658}, A127\relax
\mciteBstWouldAddEndPuncttrue
\mciteSetBstMidEndSepPunct{\mcitedefaultmidpunct}
{\mcitedefaultendpunct}{\mcitedefaultseppunct}\relax
\EndOfBibitem
\bibitem[Persson \emph{et~al.}(2018)Persson, Jørgensen, Müller, Coutens,
  Dishoeck, Taquet, Calcutt, Wiel, Bourke, and Wampfler]{per18}
M.~V. Persson, J.~K. Jørgensen, H.~S.~P. Müller, A.~Coutens, E.~F.~v.
  Dishoeck, V.~Taquet, H.~Calcutt, M.~H. D. v.~d. Wiel, T.~L. Bourke and S.~F.
  Wampfler, \emph{Astronomy \& Astrophysics}, 2018, \textbf{610}, A54\relax
\mciteBstWouldAddEndPuncttrue
\mciteSetBstMidEndSepPunct{\mcitedefaultmidpunct}
{\mcitedefaultendpunct}{\mcitedefaultseppunct}\relax
\EndOfBibitem
\bibitem[Watson(1976)]{wat76}
W.~D. Watson, \emph{Reviews of Modern Physics}, 1976, \textbf{48},
  513--552\relax
\mciteBstWouldAddEndPuncttrue
\mciteSetBstMidEndSepPunct{\mcitedefaultmidpunct}
{\mcitedefaultendpunct}{\mcitedefaultseppunct}\relax
\EndOfBibitem
\bibitem[Millar \emph{et~al.}(1989)Millar, Bennett, and Herbst]{mil89}
T.~J. Millar, A.~Bennett and E.~Herbst, \emph{The Astrophysical Journal}, 1989,
  \textbf{340}, 906\relax
\mciteBstWouldAddEndPuncttrue
\mciteSetBstMidEndSepPunct{\mcitedefaultmidpunct}
{\mcitedefaultendpunct}{\mcitedefaultseppunct}\relax
\EndOfBibitem
\bibitem[Drozdovskaya \emph{et~al.}(2019)Drozdovskaya, van Dishoeck, Rubin,
  Jørgensen, and Altwegg]{dro19}
M.~N. Drozdovskaya, E.~F. van Dishoeck, M.~Rubin, J.~K. Jørgensen and
  K.~Altwegg, \emph{Monthly Notices of the Royal Astronomical Society}, 2019,
  \textbf{490}, 50--79\relax
\mciteBstWouldAddEndPuncttrue
\mciteSetBstMidEndSepPunct{\mcitedefaultmidpunct}
{\mcitedefaultendpunct}{\mcitedefaultseppunct}\relax
\EndOfBibitem
\bibitem[Jaeger \emph{et~al.}(2018)Jaeger, Fulle, and Turk]{jae18}
S.~Jaeger, S.~Fulle and S.~Turk, \emph{Journal of Chemical Information and
  Modeling}, 2018, \textbf{58}, 27--35\relax
\mciteBstWouldAddEndPuncttrue
\mciteSetBstMidEndSepPunct{\mcitedefaultmidpunct}
{\mcitedefaultendpunct}{\mcitedefaultseppunct}\relax
\EndOfBibitem
\bibitem[Kim \emph{et~al.}(2021)Kim, Chen, Cheng, Gindulyte, He, He, Li,
  Shoemaker, Thiessen, Yu, Zaslavsky, Zhang, and Bolton]{pubchem}
S.~Kim, J.~Chen, T.~Cheng, A.~Gindulyte, J.~He, S.~He, Q.~Li, B.~A. Shoemaker,
  P.~A. Thiessen, B.~Yu, L.~Zaslavsky, J.~Zhang and E.~E. Bolton, \emph{Nucleic
  Acids Research}, 2021, \textbf{49}, D1388--D1395\relax
\mciteBstWouldAddEndPuncttrue
\mciteSetBstMidEndSepPunct{\mcitedefaultmidpunct}
{\mcitedefaultendpunct}{\mcitedefaultseppunct}\relax
\EndOfBibitem
\bibitem[Sterling and Irwin(2015)]{zinc}
T.~Sterling and J.~J. Irwin, \emph{Journal of Chemical Information and
  Modeling}, 2015, \textbf{55}, 2324--2337\relax
\mciteBstWouldAddEndPuncttrue
\mciteSetBstMidEndSepPunct{\mcitedefaultmidpunct}
{\mcitedefaultendpunct}{\mcitedefaultseppunct}\relax
\EndOfBibitem
\bibitem[Boersma \emph{et~al.}(2014)Boersma, Bauschlicher, Ricca, Mattioda,
  Cami, Peeters, de~Armas, Saborido, Hudgins, and Allamandola]{nasa1}
C.~Boersma, C.~W. Bauschlicher, A.~Ricca, A.~L. Mattioda, J.~Cami, E.~Peeters,
  F.~S. de~Armas, G.~P. Saborido, D.~M. Hudgins and L.~J. Allamandola,
  \emph{The Astrophysical Journal Supplement Series}, 2014, \textbf{211},
  8\relax
\mciteBstWouldAddEndPuncttrue
\mciteSetBstMidEndSepPunct{\mcitedefaultmidpunct}
{\mcitedefaultendpunct}{\mcitedefaultseppunct}\relax
\EndOfBibitem
\bibitem[Bauschlicher \emph{et~al.}(2018)Bauschlicher, Ricca, Boersma, and
  Allamandola]{nasa2}
C.~W. Bauschlicher, A.~Ricca, C.~Boersma and L.~J. Allamandola, \emph{The
  Astrophysical Journal Supplement Series}, 2018, \textbf{234}, 32\relax
\mciteBstWouldAddEndPuncttrue
\mciteSetBstMidEndSepPunct{\mcitedefaultmidpunct}
{\mcitedefaultendpunct}{\mcitedefaultseppunct}\relax
\EndOfBibitem
\bibitem[Mattioda \emph{et~al.}(2020)Mattioda, Hudgins, Boersma, Bauschlicher,
  Ricca, Cami, Peeters, de~Armas, Saborido, and Allamandola]{nasa3}
A.~L. Mattioda, D.~M. Hudgins, C.~Boersma, C.~W. Bauschlicher, A.~Ricca,
  J.~Cami, E.~Peeters, F.~S. de~Armas, G.~P. Saborido and L.~J. Allamandola,
  \emph{The Astrophysical Journal Supplement Series}, 2020, \textbf{251},
  22\relax
\mciteBstWouldAddEndPuncttrue
\mciteSetBstMidEndSepPunct{\mcitedefaultmidpunct}
{\mcitedefaultendpunct}{\mcitedefaultseppunct}\relax
\EndOfBibitem
\bibitem[Morgan(1965)]{mor65}
H.~L. Morgan, \emph{Journal of Chemical Documentation}, 1965, \textbf{5},
  107--113\relax
\mciteBstWouldAddEndPuncttrue
\mciteSetBstMidEndSepPunct{\mcitedefaultmidpunct}
{\mcitedefaultendpunct}{\mcitedefaultseppunct}\relax
\EndOfBibitem
\bibitem[RDKit, online()]{rdkit}
\emph{{RDK}it: Open-source cheminformatics}, \url{http://www.rdkit.org}\relax
\mciteBstWouldAddEndPuncttrue
\mciteSetBstMidEndSepPunct{\mcitedefaultmidpunct}
{\mcitedefaultendpunct}{\mcitedefaultseppunct}\relax
\EndOfBibitem
\bibitem[Rasmussen and Williams(2005)]{gpr}
C.~E. Rasmussen and C.~K.~I. Williams, \emph{Gaussian {Processes} for {Machine}
  {Learning}}, 2005\relax
\mciteBstWouldAddEndPuncttrue
\mciteSetBstMidEndSepPunct{\mcitedefaultmidpunct}
{\mcitedefaultendpunct}{\mcitedefaultseppunct}\relax
\EndOfBibitem
\bibitem[Pedregosa \emph{et~al.}(2011)Pedregosa, Varoquaux, Gramfort, Michel,
  Thirion, Grisel, Blondel, Prettenhofer, Weiss, Dubourg, Vanderplas, Passos,
  Cournapeau, Brucher, Perrot, and Duchesnay]{scikit-learn}
F.~Pedregosa, G.~Varoquaux, A.~Gramfort, V.~Michel, B.~Thirion, O.~Grisel,
  M.~Blondel, P.~Prettenhofer, R.~Weiss, V.~Dubourg, J.~Vanderplas, A.~Passos,
  D.~Cournapeau, M.~Brucher, M.~Perrot and E.~Duchesnay, \emph{Journal of
  Machine Learning Research}, 2011, \textbf{12}, 2825--2830\relax
\mciteBstWouldAddEndPuncttrue
\mciteSetBstMidEndSepPunct{\mcitedefaultmidpunct}
{\mcitedefaultendpunct}{\mcitedefaultseppunct}\relax
\EndOfBibitem
\bibitem[McGuire(2022)]{mcg21}
B.~A. McGuire, \emph{The Astrophysical Journal Supplement Series}, 2022,
  \textbf{259}, 30\relax
\mciteBstWouldAddEndPuncttrue
\mciteSetBstMidEndSepPunct{\mcitedefaultmidpunct}
{\mcitedefaultendpunct}{\mcitedefaultseppunct}\relax
\EndOfBibitem
\bibitem[Linnartz \emph{et~al.}(2015)Linnartz, Ioppolo, and Fedoseev]{lin15}
H.~Linnartz, S.~Ioppolo and G.~Fedoseev, \emph{International Reviews in
  Physical Chemistry}, 2015, \textbf{34}, 205--237\relax
\mciteBstWouldAddEndPuncttrue
\mciteSetBstMidEndSepPunct{\mcitedefaultmidpunct}
{\mcitedefaultendpunct}{\mcitedefaultseppunct}\relax
\EndOfBibitem
\bibitem[Fedoseev \emph{et~al.}(2015)Fedoseev, Cuppen, Ioppolo, Lamberts, and
  Linnartz]{fed15}
G.~Fedoseev, H.~M. Cuppen, S.~Ioppolo, T.~Lamberts and H.~Linnartz,
  \emph{Monthly Notices of the Royal Astronomical Society}, 2015, \textbf{448},
  1288--1297\relax
\mciteBstWouldAddEndPuncttrue
\mciteSetBstMidEndSepPunct{\mcitedefaultmidpunct}
{\mcitedefaultendpunct}{\mcitedefaultseppunct}\relax
\EndOfBibitem
\bibitem[Woon(2002)]{woo02}
D.~E. Woon, \emph{The Astrophysical Journal}, 2002, \textbf{569}, 541\relax
\mciteBstWouldAddEndPuncttrue
\mciteSetBstMidEndSepPunct{\mcitedefaultmidpunct}
{\mcitedefaultendpunct}{\mcitedefaultseppunct}\relax
\EndOfBibitem
\bibitem[Garrod \emph{et~al.}(2008)Garrod, Widicus~Weaver, and Herbst]{gar08}
R.~T. Garrod, S.~L. Widicus~Weaver and E.~Herbst, \emph{The Astrophysical
  Journal}, 2008, \textbf{682}, 283--302\relax
\mciteBstWouldAddEndPuncttrue
\mciteSetBstMidEndSepPunct{\mcitedefaultmidpunct}
{\mcitedefaultendpunct}{\mcitedefaultseppunct}\relax
\EndOfBibitem
\bibitem[Bergman \emph{et~al.}(2011)Bergman, Parise, Liseau, Larsson, Olofsson,
  Menten, and Güsten]{ber11}
P.~Bergman, B.~Parise, R.~Liseau, B.~Larsson, H.~Olofsson, K.~M. Menten and
  R.~Güsten, \emph{Astronomy \& Astrophysics}, 2011, \textbf{531}, L8\relax
\mciteBstWouldAddEndPuncttrue
\mciteSetBstMidEndSepPunct{\mcitedefaultmidpunct}
{\mcitedefaultendpunct}{\mcitedefaultseppunct}\relax
\EndOfBibitem
\bibitem[Du \emph{et~al.}(2012)Du, Parise, and Bergman]{du12}
F.~Du, B.~Parise and P.~Bergman, \emph{Astronomy \& Astrophysics}, 2012,
  \textbf{538}, A91\relax
\mciteBstWouldAddEndPuncttrue
\mciteSetBstMidEndSepPunct{\mcitedefaultmidpunct}
{\mcitedefaultendpunct}{\mcitedefaultseppunct}\relax
\EndOfBibitem
\bibitem[Parise \emph{et~al.}(2012)Parise, Bergman, and Du]{par12}
B.~Parise, P.~Bergman and F.~Du, \emph{Astronomy \& Astrophysics}, 2012,
  \textbf{541}, L11\relax
\mciteBstWouldAddEndPuncttrue
\mciteSetBstMidEndSepPunct{\mcitedefaultmidpunct}
{\mcitedefaultendpunct}{\mcitedefaultseppunct}\relax
\EndOfBibitem
\bibitem[Thelen \emph{et~al.}(1993)Thelen, Felder, and Robert~Huber]{the93}
M.~A. Thelen, P.~Felder and J.~Robert~Huber, \emph{Chemical Physics Letters},
  1993, \textbf{213}, 275--281\relax
\mciteBstWouldAddEndPuncttrue
\mciteSetBstMidEndSepPunct{\mcitedefaultmidpunct}
{\mcitedefaultendpunct}{\mcitedefaultseppunct}\relax
\EndOfBibitem
\bibitem[McNally(1965)]{mcn65}
D.~McNally, \emph{Science Progress (1933- )}, 1965, \textbf{53}, 83--87\relax
\mciteBstWouldAddEndPuncttrue
\mciteSetBstMidEndSepPunct{\mcitedefaultmidpunct}
{\mcitedefaultendpunct}{\mcitedefaultseppunct}\relax
\EndOfBibitem
\bibitem[Cernicharo \emph{et~al.}(2012)Cernicharo, Marcelino, Roueff, Gerin,
  Jiménez-Escobar, and Caro]{cer12}
J.~Cernicharo, N.~Marcelino, E.~Roueff, M.~Gerin, A.~Jiménez-Escobar and
  G.~M.~M. Caro, \emph{The Astrophysical Journal Letters}, 2012, \textbf{759},
  L43\relax
\mciteBstWouldAddEndPuncttrue
\mciteSetBstMidEndSepPunct{\mcitedefaultmidpunct}
{\mcitedefaultendpunct}{\mcitedefaultseppunct}\relax
\EndOfBibitem
\bibitem[Tyblewski \emph{et~al.}(1992)Tyblewski, Ha, Meyer, Bauder, and
  Blom]{tyb92}
M.~Tyblewski, T.~Ha, R.~Meyer, A.~Bauder and C.~E. Blom, \emph{The Journal of
  Chemical Physics}, 1992, \textbf{97}, 6168--6180\relax
\mciteBstWouldAddEndPuncttrue
\mciteSetBstMidEndSepPunct{\mcitedefaultmidpunct}
{\mcitedefaultendpunct}{\mcitedefaultseppunct}\relax
\EndOfBibitem
\bibitem[Möhlmann(1987)]{moh87}
G.~R. Möhlmann, \emph{Journal of Raman Spectroscopy}, 1987, \textbf{18},
  199--203\relax
\mciteBstWouldAddEndPuncttrue
\mciteSetBstMidEndSepPunct{\mcitedefaultmidpunct}
{\mcitedefaultendpunct}{\mcitedefaultseppunct}\relax
\EndOfBibitem
\bibitem[Matsuura \emph{et~al.}(1980)Matsuura, Yamamoto, and Murata]{mat80}
H.~Matsuura, M.~Yamamoto and H.~Murata, \emph{Spectrochimica Acta Part A:
  Molecular Spectroscopy}, 1980, \textbf{36}, 321--327\relax
\mciteBstWouldAddEndPuncttrue
\mciteSetBstMidEndSepPunct{\mcitedefaultmidpunct}
{\mcitedefaultendpunct}{\mcitedefaultseppunct}\relax
\EndOfBibitem
\bibitem[Ryabova \emph{et~al.}(2002)Ryabova, Voloshenko, Maiorov, and
  Osipova]{rya02}
R.~S. Ryabova, G.~I. Voloshenko, V.~D. Maiorov and G.~F. Osipova, \emph{Russian
  Journal of Applied Chemistry}, 2002, \textbf{75}, 22--24\relax
\mciteBstWouldAddEndPuncttrue
\mciteSetBstMidEndSepPunct{\mcitedefaultmidpunct}
{\mcitedefaultendpunct}{\mcitedefaultseppunct}\relax
\EndOfBibitem
\bibitem[Zhu \emph{et~al.}(2022)Zhu, Kleimeier, Turner, Singh, Fortenberry, and
  Kaiser]{zhu22}
C.~Zhu, N.~F. Kleimeier, A.~M. Turner, S.~K. Singh, R.~C. Fortenberry and R.~I.
  Kaiser, \emph{Proceedings of the National Academy of Sciences}, 2022,
  \textbf{119}, e2111938119\relax
\mciteBstWouldAddEndPuncttrue
\mciteSetBstMidEndSepPunct{\mcitedefaultmidpunct}
{\mcitedefaultendpunct}{\mcitedefaultseppunct}\relax
\EndOfBibitem
\bibitem[Manigand \emph{et~al.}(2020)Manigand, Jørgensen, Calcutt, Müller,
  Ligterink, Coutens, Drozdovskaya, Dishoeck, and Wampfler]{man20}
S.~Manigand, J.~K. Jørgensen, H.~Calcutt, H.~S.~P. Müller, N.~F.~W.
  Ligterink, A.~Coutens, M.~N. Drozdovskaya, E.~F.~v. Dishoeck and S.~F.
  Wampfler, \emph{Astronomy \& Astrophysics}, 2020, \textbf{635}, A48\relax
\mciteBstWouldAddEndPuncttrue
\mciteSetBstMidEndSepPunct{\mcitedefaultmidpunct}
{\mcitedefaultendpunct}{\mcitedefaultseppunct}\relax
\EndOfBibitem
\bibitem[Cernicharo \emph{et~al.}(2012)Cernicharo, Marcelino, Roueff, Gerin,
  Jiménez-Escobar, and Caro]{ch3o}
J.~Cernicharo, N.~Marcelino, E.~Roueff, M.~Gerin, A.~Jiménez-Escobar and
  G.~M.~M. Caro, \emph{The Astrophysical Journal Letters}, 2012, \textbf{759},
  L43\relax
\mciteBstWouldAddEndPuncttrue
\mciteSetBstMidEndSepPunct{\mcitedefaultmidpunct}
{\mcitedefaultendpunct}{\mcitedefaultseppunct}\relax
\EndOfBibitem
\bibitem[Buckley and Brochu(1972)]{buc72}
P.~Buckley and M.~Brochu, \emph{Canadian Journal of Chemistry}, 1972,
  \textbf{50}, 1149--1156\relax
\mciteBstWouldAddEndPuncttrue
\mciteSetBstMidEndSepPunct{\mcitedefaultmidpunct}
{\mcitedefaultendpunct}{\mcitedefaultseppunct}\relax
\EndOfBibitem
\bibitem[Mumma \emph{et~al.}(1996)Mumma, DiSanti, Russo, Fomenkova,
  Magee-Sauer, Kaminski, and Xie]{mum96}
M.~J. Mumma, M.~A. DiSanti, N.~D. Russo, M.~Fomenkova, K.~Magee-Sauer, C.~D.
  Kaminski and D.~X. Xie, \emph{Science}, 1996, \textbf{272}, 1310--1314\relax
\mciteBstWouldAddEndPuncttrue
\mciteSetBstMidEndSepPunct{\mcitedefaultmidpunct}
{\mcitedefaultendpunct}{\mcitedefaultseppunct}\relax
\EndOfBibitem
\bibitem[D'Hendecourt and Jourdain~de Muizon(1989)]{dhe_co2_89}
L.~B. D'Hendecourt and M.~Jourdain~de Muizon, \emph{Astronomy \& Astrophysics},
  1989, \textbf{223}, L5--L8\relax
\mciteBstWouldAddEndPuncttrue
\mciteSetBstMidEndSepPunct{\mcitedefaultmidpunct}
{\mcitedefaultendpunct}{\mcitedefaultseppunct}\relax
\EndOfBibitem
\bibitem[van Dishoeck \emph{et~al.}(1996)van Dishoeck, Helmich, de~Graauw,
  Black, Boogert, Ehrenfreund, Gerakines, Lacy, Millar, Schutte, Tielens,
  Whittet, Boxhoorn, Kester, Leech, Roelfsema, Salama, and
  Vandenbussche]{van_co2_96}
E.~F. van Dishoeck, F.~P. Helmich, T.~de~Graauw, J.~H. Black, A.~C.~A. Boogert,
  P.~Ehrenfreund, P.~A. Gerakines, J.~H. Lacy, T.~J. Millar, W.~A. Schutte,
  A.~G. G.~M. Tielens, D.~C.~B. Whittet, D.~R. Boxhoorn, D.~J.~M. Kester,
  K.~Leech, P.~R. Roelfsema, A.~Salama and B.~Vandenbussche, \emph{Astronomy \&
  Astrophysics}, 1996, \textbf{315}, L349--L352\relax
\mciteBstWouldAddEndPuncttrue
\mciteSetBstMidEndSepPunct{\mcitedefaultmidpunct}
{\mcitedefaultendpunct}{\mcitedefaultseppunct}\relax
\EndOfBibitem
\bibitem[Ward \emph{et~al.}(2012)Ward, Hogg, and Price]{ward_12_cs2}
M.~D. Ward, I.~A. Hogg and S.~D. Price, \emph{Monthly Notices of the Royal
  Astronomical Society}, 2012, \textbf{425}, 1264--1269\relax
\mciteBstWouldAddEndPuncttrue
\mciteSetBstMidEndSepPunct{\mcitedefaultmidpunct}
{\mcitedefaultendpunct}{\mcitedefaultseppunct}\relax
\EndOfBibitem
\bibitem[Drozdovskaya \emph{et~al.}(2018)Drozdovskaya, van Dishoeck,
  Jørgensen, Calmonte, van~der Wiel, Coutens, Calcutt, Müller, Bjerkeli,
  Persson, Wampfler, and Altwegg]{dro18}
M.~N. Drozdovskaya, E.~F. van Dishoeck, J.~K. Jørgensen, U.~Calmonte, M.~H.~D.
  van~der Wiel, A.~Coutens, H.~Calcutt, H.~S.~P. Müller, P.~Bjerkeli, M.~V.
  Persson, S.~F. Wampfler and K.~Altwegg, \emph{Monthly Notices of the Royal
  Astronomical Society}, 2018, \textbf{476}, 4949--4964\relax
\mciteBstWouldAddEndPuncttrue
\mciteSetBstMidEndSepPunct{\mcitedefaultmidpunct}
{\mcitedefaultendpunct}{\mcitedefaultseppunct}\relax
\EndOfBibitem
\bibitem[Zhou \emph{et~al.}(2020)Zhou, Quan, Zhang, and Qin]{zho20}
Y.~Zhou, D.-H. Quan, X.~Zhang and S.-L. Qin, \emph{Research in Astronomy and
  Astrophysics}, 2020, \textbf{20}, 125\relax
\mciteBstWouldAddEndPuncttrue
\mciteSetBstMidEndSepPunct{\mcitedefaultmidpunct}
{\mcitedefaultendpunct}{\mcitedefaultseppunct}\relax
\EndOfBibitem
\bibitem[Wang \emph{et~al.}(2023)Wang, Marks, Turner, Nikolayev, Azyazov,
  Mebel, and Kaiser]{wan23}
J.~Wang, J.~H. Marks, A.~M. Turner, A.~A. Nikolayev, V.~Azyazov, A.~M. Mebel
  and R.~I. Kaiser, \emph{Physical Chemistry Chemical Physics}, 2023,
  \textbf{25}, 936--953\relax
\mciteBstWouldAddEndPuncttrue
\mciteSetBstMidEndSepPunct{\mcitedefaultmidpunct}
{\mcitedefaultendpunct}{\mcitedefaultseppunct}\relax
\EndOfBibitem
\bibitem[Collaboration(2022)]{astropy}
T.~A. Collaboration, \emph{The Astrophysical Journal}, 2022, \textbf{935},
  167\relax
\mciteBstWouldAddEndPuncttrue
\mciteSetBstMidEndSepPunct{\mcitedefaultmidpunct}
{\mcitedefaultendpunct}{\mcitedefaultseppunct}\relax
\EndOfBibitem
\bibitem[Abuter \emph{et~al.}(2019)Abuter, Amorim, Bauböck, Berger, Bonnet,
  Brandner, Clénet, Foresto, Zeeuw, Dexter, Duvert, Eckart, Eisenhauer,
  Schreiber, Garcia, Gao, Gendron, Genzel, Gerhard, Gillessen, Habibi, Haubois,
  Henning, Hippler, Horrobin, Jiménez-Rosales, Jocou, Kervella, Lacour,
  Lapeyrère, Bouquin, Léna, Ott, Paumard, Perraut, Perrin, Pfuhl, Rabien,
  Coira, Rousset, Scheithauer, Sternberg, Straub, Straubmeier, Sturm, Tacconi,
  Vincent, Fellenberg, Waisberg, Widmann, Wieprecht, Wiezorrek, Woillez, and
  Yazici]{dgc}
R.~Abuter, A.~Amorim, M.~Bauböck, J.~P. Berger, H.~Bonnet, W.~Brandner,
  Y.~Clénet, V.~C.~d. Foresto, P.~T.~d. Zeeuw, J.~Dexter, G.~Duvert,
  A.~Eckart, F.~Eisenhauer, N.~M.~F. Schreiber, P.~Garcia, F.~Gao, E.~Gendron,
  R.~Genzel, O.~Gerhard, S.~Gillessen, M.~Habibi, X.~Haubois, T.~Henning,
  S.~Hippler, M.~Horrobin, A.~Jiménez-Rosales, L.~Jocou, P.~Kervella,
  S.~Lacour, V.~Lapeyrère, J.-B.~L. Bouquin, P.~Léna, T.~Ott, T.~Paumard,
  K.~Perraut, G.~Perrin, O.~Pfuhl, S.~Rabien, G.~R. Coira, G.~Rousset,
  S.~Scheithauer, A.~Sternberg, O.~Straub, C.~Straubmeier, E.~Sturm, L.~J.
  Tacconi, F.~Vincent, S.~v. Fellenberg, I.~Waisberg, F.~Widmann, E.~Wieprecht,
  E.~Wiezorrek, J.~Woillez and S.~Yazici, \emph{Astronomy \& Astrophysics},
  2019, \textbf{625}, L10\relax
\mciteBstWouldAddEndPuncttrue
\mciteSetBstMidEndSepPunct{\mcitedefaultmidpunct}
{\mcitedefaultendpunct}{\mcitedefaultseppunct}\relax
\EndOfBibitem
\bibitem[Dzib \emph{et~al.}(2018)Dzib, Ortiz-León, Hernández-Gómez, Loinard,
  Mioduszewski, Claussen, Menten, Caux, and Sanna]{iras_distance}
S.~A. Dzib, G.~N. Ortiz-León, A.~Hernández-Gómez, L.~Loinard, A.~J.
  Mioduszewski, M.~Claussen, K.~M. Menten, E.~Caux and A.~Sanna,
  \emph{Astronomy \& Astrophysics}, 2018, \textbf{614}, A20\relax
\mciteBstWouldAddEndPuncttrue
\mciteSetBstMidEndSepPunct{\mcitedefaultmidpunct}
{\mcitedefaultendpunct}{\mcitedefaultseppunct}\relax
\EndOfBibitem
\bibitem[Langer \emph{et~al.}(1984)Langer, Graedel, Frerking, and
  Armentrout]{lan84}
W.~D. Langer, T.~E. Graedel, M.~A. Frerking and P.~B. Armentrout, \emph{The
  Astrophysical Journal}, 1984, \textbf{277}, 581--604\relax
\mciteBstWouldAddEndPuncttrue
\mciteSetBstMidEndSepPunct{\mcitedefaultmidpunct}
{\mcitedefaultendpunct}{\mcitedefaultseppunct}\relax
\EndOfBibitem
\bibitem[Smith \emph{et~al.}(2021)Smith, Gudipati, Smith, and Lewis]{smi21}
L.~R. Smith, M.~S. Gudipati, R.~L. Smith and R.~D. Lewis, \emph{Astronomy \&
  Astrophysics}, 2021, \textbf{656}, A82\relax
\mciteBstWouldAddEndPuncttrue
\mciteSetBstMidEndSepPunct{\mcitedefaultmidpunct}
{\mcitedefaultendpunct}{\mcitedefaultseppunct}\relax
\EndOfBibitem
\bibitem[Jørgensen \emph{et~al.}(2011)Jørgensen, Bourke, Luong, and {S.
  Takakuwa}]{jor11_iso}
J.~K. Jørgensen, T.~L. Bourke, Q.~N. Luong and {S. Takakuwa}, \emph{Astronomy
  \& Astrophysics}, 2011, \textbf{534}, A100\relax
\mciteBstWouldAddEndPuncttrue
\mciteSetBstMidEndSepPunct{\mcitedefaultmidpunct}
{\mcitedefaultendpunct}{\mcitedefaultseppunct}\relax
\EndOfBibitem
\bibitem[Carvajal \emph{et~al.}(2009)Carvajal, Margulès, Tercero, Demyk,
  Kleiner, Guillemin, Lattanzi, Walters, Demaison, Wlodarczak, Huet,
  Møllendal, Ilyushin, and Cernicharo]{formate_iso}
M.~Carvajal, L.~Margulès, B.~Tercero, K.~Demyk, I.~Kleiner, J.~C. Guillemin,
  V.~Lattanzi, A.~Walters, J.~Demaison, G.~Wlodarczak, T.~R. Huet,
  H.~Møllendal, V.~V. Ilyushin and J.~Cernicharo, \emph{Astronomy \&
  Astrophysics}, 2009, \textbf{500}, 1109--1118\relax
\mciteBstWouldAddEndPuncttrue
\mciteSetBstMidEndSepPunct{\mcitedefaultmidpunct}
{\mcitedefaultendpunct}{\mcitedefaultseppunct}\relax
\EndOfBibitem
\bibitem[Müller \emph{et~al.}(2005)Müller, Schlöder, Stutzki, and
  Winnewisser]{cdms}
H.~S.~P. Müller, F.~Schlöder, J.~Stutzki and G.~Winnewisser, \emph{Journal of
  Molecular Structure}, 2005, \textbf{742}, 215--227\relax
\mciteBstWouldAddEndPuncttrue
\mciteSetBstMidEndSepPunct{\mcitedefaultmidpunct}
{\mcitedefaultendpunct}{\mcitedefaultseppunct}\relax
\EndOfBibitem
\bibitem[Smith \emph{et~al.}(2020)Smith, Burns, Simmonett, Parrish, Schieber,
  Galvelis, Kraus, Kruse, Di~Remigio, Alenaizan, James, Lehtola, Misiewicz,
  Scheurer, Shaw, Schriber, Xie, Glick, Sirianni, O’Brien, Waldrop, Kumar,
  Hohenstein, Pritchard, Brooks, Schaefer, Sokolov, Patkowski, DePrince,
  Bozkaya, King, Evangelista, Turney, Crawford, and Sherrill]{psi4}
D.~G.~A. Smith, L.~A. Burns, A.~C. Simmonett, R.~M. Parrish, M.~C. Schieber,
  R.~Galvelis, P.~Kraus, H.~Kruse, R.~Di~Remigio, A.~Alenaizan, A.~M. James,
  S.~Lehtola, J.~P. Misiewicz, M.~Scheurer, R.~A. Shaw, J.~B. Schriber, Y.~Xie,
  Z.~L. Glick, D.~A. Sirianni, J.~S. O’Brien, J.~M. Waldrop, A.~Kumar, E.~G.
  Hohenstein, B.~P. Pritchard, B.~R. Brooks, H.~F. Schaefer, A.~Y. Sokolov,
  K.~Patkowski, A.~E. DePrince, U.~Bozkaya, R.~A. King, F.~A. Evangelista,
  J.~M. Turney, T.~D. Crawford and C.~D. Sherrill, \emph{The Journal of
  Chemical Physics}, 2020, \textbf{152}, 184108\relax
\mciteBstWouldAddEndPuncttrue
\mciteSetBstMidEndSepPunct{\mcitedefaultmidpunct}
{\mcitedefaultendpunct}{\mcitedefaultseppunct}\relax
\EndOfBibitem
\bibitem[Pickett(1991)]{spcat}
H.~M. Pickett, \emph{Journal of Molecular Spectroscopy}, 1991, \textbf{148},
  371--377\relax
\mciteBstWouldAddEndPuncttrue
\mciteSetBstMidEndSepPunct{\mcitedefaultmidpunct}
{\mcitedefaultendpunct}{\mcitedefaultseppunct}\relax
\EndOfBibitem
\bibitem[McGuire and Lee(2020)]{molsim}
B.~A. McGuire and K.~Lee, \emph{molsim}, 2020,
  \url{https://zenodo.org/record/4122749}\relax
\mciteBstWouldAddEndPuncttrue
\mciteSetBstMidEndSepPunct{\mcitedefaultmidpunct}
{\mcitedefaultendpunct}{\mcitedefaultseppunct}\relax
\EndOfBibitem
\bibitem[Lykke \emph{et~al.}(2017)Lykke, Coutens, Jørgensen, Wiel, Garrod,
  Müller, Bjerkeli, Bourke, Calcutt, Drozdovskaya, Favre, Fayolle, Jacobsen,
  Öberg, Persson, Dishoeck, and Wampfler]{lyk17}
J.~M. Lykke, A.~Coutens, J.~K. Jørgensen, M.~H. D. v.~d. Wiel, R.~T. Garrod,
  H.~S.~P. Müller, P.~Bjerkeli, T.~L. Bourke, H.~Calcutt, M.~N. Drozdovskaya,
  C.~Favre, E.~C. Fayolle, S.~K. Jacobsen, K.~I. Öberg, M.~V. Persson,
  E.~F.~v. Dishoeck and S.~F. Wampfler, \emph{Astronomy \& Astrophysics}, 2017,
  \textbf{597}, A53\relax
\mciteBstWouldAddEndPuncttrue
\mciteSetBstMidEndSepPunct{\mcitedefaultmidpunct}
{\mcitedefaultendpunct}{\mcitedefaultseppunct}\relax
\EndOfBibitem
\bibitem[Calcutt \emph{et~al.}(2018)Calcutt, Jørgensen, Müller, Kristensen,
  Coutens, Bourke, Garrod, Persson, Wiel, Dishoeck, and Wampfler]{cal18b}
H.~Calcutt, J.~K. Jørgensen, H.~S.~P. Müller, L.~E. Kristensen, A.~Coutens,
  T.~L. Bourke, R.~T. Garrod, M.~V. Persson, M.~H. D. v.~d. Wiel, E.~F.~v.
  Dishoeck and S.~F. Wampfler, \emph{Astronomy \& Astrophysics}, 2018,
  \textbf{616}, A90\relax
\mciteBstWouldAddEndPuncttrue
\mciteSetBstMidEndSepPunct{\mcitedefaultmidpunct}
{\mcitedefaultendpunct}{\mcitedefaultseppunct}\relax
\EndOfBibitem
\bibitem[Calcutt \emph{et~al.}(2018)Calcutt, Fiechter, Willis, Müller, Garrod,
  Jørgensen, Wampfler, Bourke, Coutens, Drozdovskaya, Ligterink, and
  Kristensen]{cal18a}
H.~Calcutt, M.~R. Fiechter, E.~R. Willis, H.~S.~P. Müller, R.~T. Garrod, J.~K.
  Jørgensen, S.~F. Wampfler, T.~L. Bourke, A.~Coutens, M.~N. Drozdovskaya,
  N.~F.~W. Ligterink and L.~E. Kristensen, \emph{Astronomy \& Astrophysics},
  2018, \textbf{617}, A95\relax
\mciteBstWouldAddEndPuncttrue
\mciteSetBstMidEndSepPunct{\mcitedefaultmidpunct}
{\mcitedefaultendpunct}{\mcitedefaultseppunct}\relax
\EndOfBibitem
\bibitem[Ligterink \emph{et~al.}(2017)Ligterink, Coutens, Kofman, Müller,
  Garrod, Calcutt, Wampfler, Jørgensen, Linnartz, and van Dishoeck]{lig17}
N.~F.~W. Ligterink, A.~Coutens, V.~Kofman, H.~S.~P. Müller, R.~T. Garrod,
  H.~Calcutt, S.~F. Wampfler, J.~K. Jørgensen, H.~Linnartz and E.~F. van
  Dishoeck, \emph{Monthly Notices of the Royal Astronomical Society}, 2017,
  \textbf{469}, 2219--2229\relax
\mciteBstWouldAddEndPuncttrue
\mciteSetBstMidEndSepPunct{\mcitedefaultmidpunct}
{\mcitedefaultendpunct}{\mcitedefaultseppunct}\relax
\EndOfBibitem
\bibitem[Fayolle \emph{et~al.}(2017)Fayolle, Öberg, Jørgensen, Altwegg,
  Calcutt, Müller, Rubin, van~der Wiel, Bjerkeli, Bourke, Coutens, van
  Dishoeck, Drozdovskaya, Garrod, Ligterink, Persson, and Wampfler]{fay17}
E.~C. Fayolle, K.~I. Öberg, J.~K. Jørgensen, K.~Altwegg, H.~Calcutt, H.~S.~P.
  Müller, M.~Rubin, M.~H.~D. van~der Wiel, P.~Bjerkeli, T.~L. Bourke,
  A.~Coutens, E.~F. van Dishoeck, M.~N. Drozdovskaya, R.~T. Garrod, N.~F.~W.
  Ligterink, M.~V. Persson and S.~F. Wampfler, \emph{Nature Astronomy}, 2017,
  \textbf{1}, 703--708\relax
\mciteBstWouldAddEndPuncttrue
\mciteSetBstMidEndSepPunct{\mcitedefaultmidpunct}
{\mcitedefaultendpunct}{\mcitedefaultseppunct}\relax
\EndOfBibitem
\bibitem[Manigand \emph{et~al.}(2021)Manigand, Coutens, Loison, Wakelam,
  Calcutt, Müller, Jørgensen, Taquet, Wampfler, Bourke, Kulterer, Dishoeck,
  Drozdovskaya, and Ligterink]{man21}
S.~Manigand, A.~Coutens, J.-C. Loison, V.~Wakelam, H.~Calcutt, H.~S.~P.
  Müller, J.~K. Jørgensen, V.~Taquet, S.~F. Wampfler, T.~L. Bourke, B.~M.
  Kulterer, E.~F.~v. Dishoeck, M.~N. Drozdovskaya and N.~F.~W. Ligterink,
  \emph{Astronomy \& Astrophysics}, 2021, \textbf{645}, A53\relax
\mciteBstWouldAddEndPuncttrue
\mciteSetBstMidEndSepPunct{\mcitedefaultmidpunct}
{\mcitedefaultendpunct}{\mcitedefaultseppunct}\relax
\EndOfBibitem
\bibitem[Calcutt \emph{et~al.}(2019)Calcutt, Willis, Jørgensen, Bjerkeli,
  Ligterink, Coutens, Müller, Garrod, Wampfler, and Drozdovskaya]{cal19}
H.~Calcutt, E.~R. Willis, J.~K. Jørgensen, P.~Bjerkeli, N.~F.~W. Ligterink,
  A.~Coutens, H.~S.~P. Müller, R.~T. Garrod, S.~F. Wampfler and M.~N.
  Drozdovskaya, \emph{Astronomy \& Astrophysics}, 2019, \textbf{631},
  A137\relax
\mciteBstWouldAddEndPuncttrue
\mciteSetBstMidEndSepPunct{\mcitedefaultmidpunct}
{\mcitedefaultendpunct}{\mcitedefaultseppunct}\relax
\EndOfBibitem
\bibitem[Coutens \emph{et~al.}(2022)Coutens, Loison, Boulanger, Caux, Müller,
  Wakelam, Manigand, and Jørgensen]{cou22}
A.~Coutens, J.-C. Loison, A.~Boulanger, E.~Caux, H.~S.~P. Müller, V.~Wakelam,
  S.~Manigand and J.~K. Jørgensen, \emph{Astronomy \& Astrophysics}, 2022,
  \textbf{660}, L6\relax
\mciteBstWouldAddEndPuncttrue
\mciteSetBstMidEndSepPunct{\mcitedefaultmidpunct}
{\mcitedefaultendpunct}{\mcitedefaultseppunct}\relax
\EndOfBibitem
\bibitem[Coutens \emph{et~al.}(2019)Coutens, Ligterink, Loison, Wakelam,
  Calcutt, Drozdovskaya, Jørgensen, Müller, Dishoeck, and Wampfler]{cou19}
A.~Coutens, N.~F.~W. Ligterink, J.-C. Loison, V.~Wakelam, H.~Calcutt, M.~N.
  Drozdovskaya, J.~K. Jørgensen, H.~S.~P. Müller, E.~F.~v. Dishoeck and S.~F.
  Wampfler, \emph{Astronomy \& Astrophysics}, 2019, \textbf{623}, L13\relax
\mciteBstWouldAddEndPuncttrue
\mciteSetBstMidEndSepPunct{\mcitedefaultmidpunct}
{\mcitedefaultendpunct}{\mcitedefaultseppunct}\relax
\EndOfBibitem
\bibitem[Ligterink \emph{et~al.}(2018)Ligterink, Calcutt, Coutens, Kristensen,
  Bourke, Drozdovskaya, Müller, Wampfler, Wiel, Dishoeck, and
  Jørgensen]{lig17b}
N.~F.~W. Ligterink, H.~Calcutt, A.~Coutens, L.~E. Kristensen, T.~L. Bourke,
  M.~N. Drozdovskaya, H.~S.~P. Müller, S.~F. Wampfler, M.~H. D. v.~d. Wiel,
  E.~F.~v. Dishoeck and J.~K. Jørgensen, \emph{Astronomy \& Astrophysics},
  2018, \textbf{619}, A28\relax
\mciteBstWouldAddEndPuncttrue
\mciteSetBstMidEndSepPunct{\mcitedefaultmidpunct}
{\mcitedefaultendpunct}{\mcitedefaultseppunct}\relax
\EndOfBibitem
\bibitem[Ligterink \emph{et~al.}(2018)Ligterink, Terwisscha van Scheltinga,
  Taquet, Jørgensen, Cazaux, van Dishoeck, and Linnartz]{acetamide}
N.~F.~W. Ligterink, J.~Terwisscha van Scheltinga, V.~Taquet, J.~K.
  Jørgensen, S.~Cazaux, E.~F. van Dishoeck and H.~Linnartz, \emph{Monthly
  Notices of the Royal Astronomical Society}, 2018, \textbf{480},
  3628--3643\relax
\mciteBstWouldAddEndPuncttrue
\mciteSetBstMidEndSepPunct{\mcitedefaultmidpunct}
{\mcitedefaultendpunct}{\mcitedefaultseppunct}\relax
\EndOfBibitem
\bibitem[Manigand \emph{et~al.}(2019)Manigand, Calcutt, Jørgensen, Taquet,
  Müller, Coutens, Wampfler, Ligterink, Drozdovskaya, Kristensen, Wiel, and
  Bourke]{man18}
S.~Manigand, H.~Calcutt, J.~K. Jørgensen, V.~Taquet, H.~S.~P. Müller,
  A.~Coutens, S.~F. Wampfler, N.~F.~W. Ligterink, M.~N. Drozdovskaya, L.~E.
  Kristensen, M.~H. D. v.~d. Wiel and T.~L. Bourke, \emph{Astronomy \&
  Astrophysics}, 2019, \textbf{623}, A69\relax
\mciteBstWouldAddEndPuncttrue
\mciteSetBstMidEndSepPunct{\mcitedefaultmidpunct}
{\mcitedefaultendpunct}{\mcitedefaultseppunct}\relax
\EndOfBibitem
\end{mcitethebibliography}

\clearpage

\section{Appendix}
\label{sec: appendix_a}

\begin{figure*}
\begin{center}
\includegraphics[width=\textwidth]{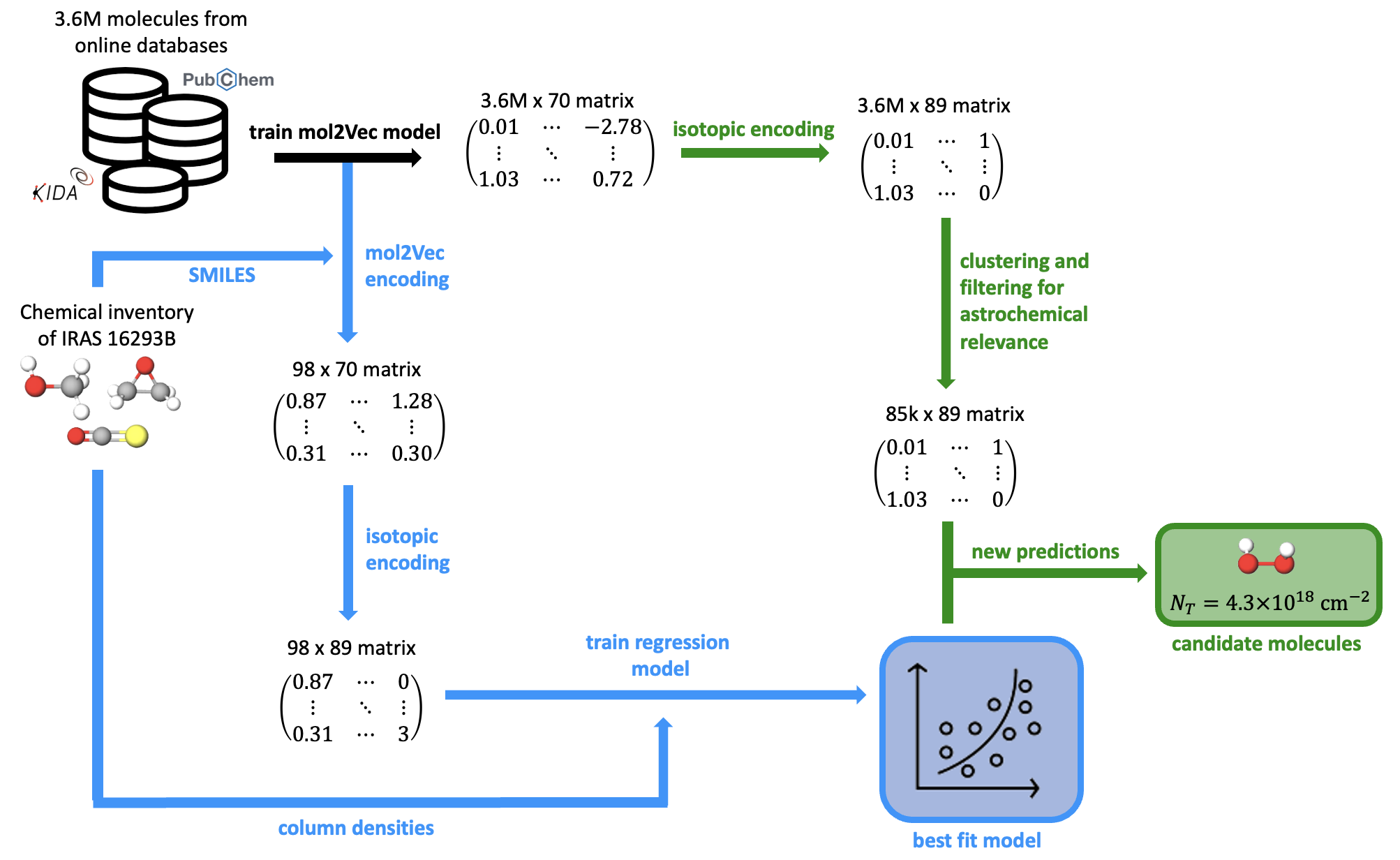}
\caption{Schematic that summarizes each of the steps of the machine learning method.}\label{schematic}
\end{center}
\end{figure*}

\begin{table*}[h!]
    \centering
    \caption{Molecules detected at a one-beam or half-beam offset position from the continuum peak of IRAS 16293B in the south-west direction. The units of the observed column densities are $\log_{10}$ \si{\per\centi\meter\squared}}. \label{appendixTable}
    \begin{tabular}{cccc}
    \toprule
    Formula &  SMILES    &   Observed Column Density &   Reference \\
    \midrule
\ce{CO} & [C-]\#[O+] & 20.0000 & \citet{dro19} \\
\ce{CH3OH}  &  CO  & 19.0000 &  \citet{jor16, jor18} \\
\ce{H2CO}  &  C=O  & 18.2800 &  \citet{per18} \\
\ce{CH3CH2OH}  &  CCO  & 17.3600 &  \citet{jor18} \\
\ce{CH3OCH3}  &  COC  & 17.3800 &  \citet{jor18} \\
\ce{HCOOCH3}  &  COC=O  & 17.4100 &  \citet{jor18} \\
\ce{CH2OHCHO}  &  O=CCO  & 16.5100 &  \citet{jor16} \\
\ce{CH3COOH}  &  CC(=O)O  & 15.4500 &  \citet{jor16} \\
\ce{CH3CHO}  &  CC=O  & 17.0800 &  \citet{jor18} \\
\ce{c-C2H4O}  &  C1CO1  & 15.7300 &  \citet{lyk17} \\
\ce{HCOOH}  &  O=CO  & 16.7500 &  \citet{jor18} \\
aGg'-\ce{(CH2OH)2}  &  OCCO  & 16.7200 &  \citet{jor16} \\
\ce{CH3OCH2OH}  &  COCO  & 17.1500 &  \citet{man20} \\
\ce{C2H5CHO}  &  CCC=O  & 15.3400 &  \citet{lyk17} \\
\ce{(CH3)2CO}  &  CC(C)=O  & 16.2300 &  \citet{lyk17} \\
\ce{NH2CHO}  &  NC=O  & 15.9800 &  \citet{cou16} \\
\ce{HCN}  &  C\#N  & 16.7000 &  \citet{dro19} \\
\ce{CH3CN}  &  CC\#N  & 16.6000 &  \citet{cal18b} \\
\ce{CH3NC}  &  [C-]\#[N+]C  & 14.3000 &  \citet{cal18a} \\
\ce{HNCO}  &  N=C=O  & 16.5700 &  \citet{lig17} \\
\ce{HC3N}  &  C\#CC\#N  & 14.2600 &  \citet{cal18b} \\
\ce{H2S}  &  S  & 17.2300 &  \citet{dro18} \\
\ce{OCS}  &  O=C=S  & 17.4000 &  \citet{dro18} \\
\ce{CH3SH}  &  CS  & 15.6800 &  \citet{dro18} \\
\ce{CS}  &  [C-]\#[S+]  & 15.5900 &  \citet{dro18} \\
\ce{H2CS}  &  C=S  & 15.1100 &  \citet{dro18} \\
\ce{SO}  &  O=S  & 14.6400 &  \citet{dro18} \\
\ce{CH3Cl}  &  CCl  & 14.6600 &  \citet{fay17} \\
\ce{C2H3CHO}  &  C=CC=O  & 14.5300 &  \citet{man21} \\
\ce{C3H6}  &  C=CC  & 16.6200 &  \citet{man21} \\
\ce{CH3CCH}  &  C\#CC  & 16.0400 &  \citet{cal19} \\
\ce{t-C2H5OCH3}  &  CCOC  & 16.2600 &  \citet{man20} \\
\ce{C3H4O2}  &  O=CC=CO  & 15.0000 &  \citet{cou22} \\
\ce{CH3NCO}  &  CN=C=O  & 15.6000 &  \citet{lig17} \\
\ce{C2H5CN}  &  CCC\#N  & 15.5600 &  \citet{cal18b} \\
\ce{C2H3CN}  &  C=CC\#N  & 14.8700 &  \citet{cal18b} \\
\ce{CH2CO}  &  C=C=O  & 16.6800 &  \citet{jor18} \\
\ce{HONO}  &  O=NO  & 14.9500 &  \citet{cou19} \\
\ce{NO}  &  [N]=O  & 16.3000 &  \citet{lig17b} \\
\ce{CH3C(O)NH2}  &  CC(N)=O  & 14.9500 &  \citet{acetamide} \\
\ce{SO2}  &  O=S=O  & 15.1100 &  \citet{dro18} \\
\ce{t-HCOOH}  &  O=CO  & 16.7500 &  \citet{jor18} \\ 
\ce{CH2NH}  &  C=N  & 14.9031 &  \citet{lig17b} \\

        \bottomrule
    \end{tabular}

    \label{table2} 
\end{table*}

\begin{table*}[h!]
    \centering
    \caption{Continued ...} \label{appendixTable}
    \begin{tabular}{cccc}
    \toprule
    Formula &   SMILES    &   Observed Column Density &   Reference \\
    \midrule
\ce{H2}\ce{^{13}CO} &  [13CH2]=O  & 16.5563 &  \citet{per18} \\
\ce{H2C^{17}O}  &  C=[17O]  & 14.8573 &  \citet{per18} \\
\ce{H2C^{18}O}  &  C=[18O]  & 15.3617 &  \citet{per18} \\
\ce{HDCO}  &  [2H]C=O  & 17.1139 &  \citet{per18} \\
\ce{D2CO}  &  [2H]C([2H])=O  & 16.2041 &  \citet{per18} \\
\ce{D2}\ce{^{13}CO}  &  [2H][13C]([2H])=O  & 14.3424 &  \citet{per18} \\
\ce{HC^{15}N}  &  C\#[15N]  & 14.3979 &  \citet{dro19} \\
\ce{^{13}CH3CN}  &  [13CH3]C\#N  & 14.7782 &  \citet{cal18b} \\
\ce{CH3}\ce{^{13}CN}  &  C[13C]\#N  & 14.6990 &  \citet{cal18b} \\
\ce{CH3C^{15}N}  &  CC\#[15N]  & 14.2041 &  \citet{cal18b} \\
\ce{CH2DCN}  &  [2H]CC\#N  & 15.1461 &  \citet{cal18b} \\
\ce{CHD2CN}  &  [2H]C([2H])C\#N  & 14.3010 &  \citet{cal18b} \\
\ce{^{34}SO2}  &  O=[34S]=O  & 14.6021 &  \citet{dro18} \\
\ce{O^{13}CS}  &  O=[13C]=S  & 15.6990 &  \citet{dro18} \\
\ce{OC^{34}S}  &  O=C=[34S]  & 16.0000 &  \citet{dro18} \\
\ce{OC^{33}S}  &  O=C=[33S]  & 15.4771 &  \citet{ dro18} \\
\ce{^{18}OCS}  &  [18O]=C=S  & 14.6990 &  \citet{dro18} \\
\ce{C^{34}S}  &  [C-]\#[34S+]  & 14.3010 &  \citet{dro18} \\
\ce{C^{33}S}  &  [C-]\#[33S+]  & 13.9031 &  \citet{dro18} \\
\ce{C^{36}S}  &  [C-]\#[36S+]  & 13.1461 &  \citet{dro18} \\
\ce{HDCS}  &  [2H]C=S  & 14.1761 &  \citet{dro18} \\
\ce{HDS}  &  [2H]S  & 16.2041 &  \citet{dro18} \\
\ce{HD^{34}S}  &  [2H][34SH]  & 15.0000 &  \citet{dro18} \\
\ce{CD3OH}  &  [2H]C([2H])([2H])O  & 16.4914 &  \citet{ily22} \\
\ce{CH2DOH}  &  [2H]CO  & 17.8513 &  \citet{jor18} \\
\ce{CH3OD}  &  [2H]OC  & 17.2553 &  \citet{jor18} \\
\ce{a-CH3CHDOH}  &  [2H]C(C)O  & 16.3617 &  \citet{jor18} \\
\ce{CH3OCDO}  &  [2H]C(=O)OC  & 16.1761 &  \citet{jor18} \\
\ce{CH2DOCHO}  &  [2H]COC=O  & 16.6812 &  \citet{jor18} \\
\ce{CHDCO}  &  [2H]C=C=O  & 15.3010 &  \citet{jor18} \\
\ce{^{13}CH3OCH3}  &  CO[13CH3]  & 16.1461 &  \citet{jor18} \\
\ce{CH3CDO}  &  [2H]C(C)=O  & 15.9823 &  \citet{jor18} \\
\ce{H^{13}COOH}  &  O=[13CH]O  & 14.9191 &  \citet{jor18} \\
\ce{CHD2OCHO}  &  [2H]C([2H])OC=O  & 16.0414 &  \citet{man18} \\
\ce{CH3}\ce{^{37}Cl}  &  C[37Cl]  & 14.3424 &  \citet{fay17} \\
\ce{NH2CDO}  &  [2H]C(N)=O  & 14.3222 &  \citet{cou16} \\
\ce{NH2}\ce{^{13}CHO}  &  N[13CH]=O  & 14.1761 &  \citet{cou16} \\
\ce{DNCO}  &  [2H]N=C=O  & 14.4771 &  \citet{cou16} \\
\ce{HN^{13C}O}  &  N=[13C]=O  & 14.6021 &  \citet{cou16} \\
\ce{CHDOHCHO}  &  [2H]C(O)C=O  & 15.5211 &  \citet{jor16} \\
\ce{CH2ODCHO}  &  [2H]OCC=O  & 15.1761 &  \citet{jor16} \\
\ce{CH2OHCDO}  &  [2H]C(=O)CO  & 15.2148 &  \citet{jor16} \\
\ce{CH3}\ce{^{18}OH}  &  C[18OH]  & 16.2718 &  \citet{jor16} \\
\ce{^{13}CH2CO}  &  [13CH2]=C=O  & 14.8513 &  \citet{jor18} \\
\ce{CH2}\ce{^{13}CO}  &  C=[13C]=O  & 14.8513 &  \citet{jor18} \\
\ce{^{13}CH3CHO}  &  [13CH3]C=O  & 15.2553 &  \citet{jor18} \\
\ce{CH3}\ce{^{13}CHO}  &  C[13CH]=O  & 15.2553 &  \citet{jor18} \\
\ce{t-DCOOH}  &  [2H]C(=O)O  & 15.0414 &  \citet{jor18} \\
\ce{t-HCOOD}  &  [2H]OC=O  & 15.0414 &  \citet{jor18} \\
\ce{a-a-CH2DCH2OH} & [2H]CCO & 16.4313 & \citet{jor18} \\
\ce{asym-CH2DOCH3} & [2H]COC & 16.6128 & \citet{jor18} \\

        \bottomrule
    \end{tabular}

    \label{table2}
\end{table*}

\balance

\end{document}